\documentclass[11pt,a4paper]{article}
\usepackage{jcappub}
\usepackage{ifthen}
\usepackage{epsfig}

\def\eps{{\cal E}}

\newcommand{\erf}{\mathop{\mathrm{erf}}}

\newcommand{\vmin}{\ensuremath{v_\mathrm{min}}}

\newcommand{\bvobs}{\ensuremath{\mathbf{v}_\mathrm{obs}}}
\newcommand{\vesc}{\ensuremath{v_\mathrm{esc}}}
\newcommand{\Nesc}{\ensuremath{N_\mathrm{esc}}}
\newcommand{\bu}{\ensuremath{\mathbf{u}}}  % bold u (velocity vector)
\newcommand{\bv}{\ensuremath{\mathbf{v}}}  % bold v (velocity vector)
\newcommand{\bV}{\ensuremath{\mathbf{V}}}  % bold V (velocity vector)

\newcommand{\eone}{\ensuremath{\hat{\mathbf{\varepsilon}}_1}}  % e_1 unit vector
\newcommand{\etwo}{\ensuremath{\hat{\mathbf{\varepsilon}}_2}}  % e_2 unit vector
\newcommand{\beqra}{\begin{eqnarray}}
\newcommand{\eeqra}{\end{eqnarray}}
\newcommand{\beq}{\begin{equation}}
\newcommand{\eeq}{\end{equation}}

\newcommand{\ds}{{\sffamily DarkSUSY}}

\title{The local dark matter phase-space density and impact on WIMP direct detection}
\author[a]{Riccardo Catena} 
\author[b]{and Piero Ullio} 
\affiliation[a]{Institut f\"ur Theoretische Physik, Friedrich-Hund-Platz 1, 37077 G\"ottingen, Germany}
\affiliation[b]{SISSA and INFN, Sezione di Trieste, Via Bonomea, 265, 34136 Trieste, Italy}
\emailAdd{riccardo.catena@theorie.physik.uni-goettingen.de}
\emailAdd{ullio@sissa.it}
\abstract{
We present a new determination of the local dark matter phase-space density. This result is obtained implementing, in the limit of isotropic velocity distribution and spherical symmetry, Eddington's inversion
formula, which links univocally the dark matter distribution function to the density profile, and applying, within 
a Bayesian framework, a Markov Chain Monte Carlo algorithm to sample mass models for the Milky Way
against a broad and variegated sample of dynamical constraints. We consider three possible choices 
for the dark matter density profile, namely the Einasto, NFW and Burkert profiles, finding that the velocity 
dispersion, which characterizes the width in the distribution, tends to be larger for the Burkert case, while
the escape velocity depends very weakly on the profile, with the mean value we obtain being in very good
agreement with estimates from stellar kinematics. The derived dark matter phase-space densities differ
significantly--most dramatically in the high velocity tails--from the model usually taken as a reference in 
dark matter detection studies, a Maxwell-Boltzmann distribution with velocity dispersion fixed in terms 
of the local circular velocity and with a sharp truncation at a given value of the escape velocity. We 
discuss the impact of astrophysical uncertainties on dark matter scattering rates and direct detection 
exclusion limits, considering a few sample cases and showing that the most sensitive ones are those 
for light dark matter particles and for particles scattering inelastically. As a general trend, regardless 
of the assumed profile, when adopting a self-consistent phase-space density, we find that rates are
larger, and hence exclusion limits stronger, than with the standard Maxwell-Boltzmann approximation.
Tools for applying our result on the local dark matter phase-space density to other dark matter 
candidates or experimental setups are provided.}

\keywords{dark matter theory, dark matter experiments, rotation curves of galaxies, dark matter simulations} 
\begin{document}
\maketitle

\section{Introduction}

The direct detection technique has played a major role in the quest for the identification of the 
dark matter (DM) component of the Universe. The goal is measure the recoil energy due to the scatterings 
on a target material of the particles forming the Milky Way DM halo. The prime focus is on weakly 
interacting massive particles (WIMPs), non-relativistic thermal relics from the Universe with mass in the 
range, say, between few GeV and few TeV, which would scatter elastically on a nucleus, depositing 
energies of the order of few (tens of) keV (for a recent review on the DM problem and direct detection
see, e.g.,~\cite{Bertone:2010zz}). In recent years several experiments have reached the sensitivity
to start probing the level of scattering cross sections expected for WIMPs. In particular, three collaborations have 
published results compatible with a positive signal: DAMA and DAMA/LIBRA~\cite{Bernabei:2008yi} detected 
an annual modulation in the total event rate consistent with the effect expected from WIMPs 
scatterings because of the orbit of the Earth around the Sun;
CoGeNT has recently confirmed~\cite{Aalseth:2011wp} the detection of a low-energy exponential tail 
in their count rate consistent with the shape predicted for the signal from a WIMP slightly lighter than 
10~GeV, as already found previously~\cite{Aalseth:2010vx}, showing in addition a 2.8~$\sigma$ indication 
in favor of an annual modulation effect; finally, CRESST-II~\cite{crestt} has just reported the detection
of a number events in the acceptance region of low energy nuclear recoils, rejecting the hypothesis
that they are due to background only, and showing the indication for a component of mass either around
25 or 11~GeV. On the other hand, other experiments such as CDMS~\cite{Ahmed:2009zw} 
and Xenon100~\cite{Aprile:2011hi} have not found any evidence for a DM signal and produced exclusion
plots in the plane WIMP mass versus scattering cross section which seem to disfavor the region of the 
parameter space preferred by DAMA, CoGeNT or CRESST-II (in this respect, see also the recent reanalysis of the data 
taken at an early stage by CDMS at a shallow site~\cite{Akerib:2010pv}, and the result of 
Xenon10~\cite{Angle:2009xb}).

The interpretation of a count rate or an annual modulation effect in a given experiment in terms of 
properties of DM particles (and hence obviously the comparison of results obtained by different collaborations, 
possibly, with different techniques and/or different target materials) depend on a number of assumptions implemented 
in the analysis of the data. Besides issues regarding understanding the target material, such as nuclear form 
factors or whether channeling occurs, and the performance of the detector, such as determining
the energy threshold, the quenching factors and background rejection/contamination, there are 
uncertainties related to the DM signal itself. First of all there is an uncertainty in the normalization 
of the incident DM particle flux, which scales with the local halo density, often quoted to be unknown
within a factor of 2 or so. Large uncertainties also affect the energy spectrum of the DM particles 
in the detector frame, in turn connected to their velocity distribution in the
Galactic frame and to the proper motion of the Sun/Earth system. Concerning these, the vast
majority of the analyses adopt a standard paradigm in which the velocity distribution is assumed
to be Maxwell-Boltzmann with velocity dispersion scaled up of a factor of $\sqrt{3/2}$ with respect to
the Sun circular velocity (usually taken as the IAU standard value of 220~km~s$^{-1}$) and 
truncated to the value assumed for the escape velocity.

The Maxwellian distribution is the configuration maximizing the entropy for a  
self-gravitating collisionless system (complete wash out of the initial conditions after gravitational 
collapse) and is associated to the spherical isothermal sphere density profile, which declines as $r^{-2}$ 
at large radii and hence supports a flat rotation curve. It is a well-motivated form but
unlikely a fair description of the Milky Way DM halo. In particular cosmological N-body simulations 
find that DM halos have density profiles falling more rapidly at large radii, as $r^{-3}$, and velocity 
distributions showing significant departures from the Maxwell-Boltzmann shape, see, e.g., the results from 
the high-resolution simulations Via Lactea~\cite{Diemand:2008in} and Aquarius~\cite{Springel:2008cc}.
A few analyses have discussed the impact on direct detection results when using 
DM phase-space distribution function as directly read out from the numerical simulations or from
other specialized approaches devised to describe in detail the fine-grained structure
of the DM velocity distribution, see, e.g., \cite{Moore:2001vq,Helmi:2002ss,Stiff:2003tx,Hansen:2005yj,
Vogelsberger:2008qb,Kuhlen:2009vh,Fantin:2008ur,Vogelsberger:2010gd,Fantin:2011nt,Lisanti:2010qx}.
The main shortcoming of these approaches is that, since it is highly non-trivial to include baryons
and baryonic feedback in the simulations, these analyses treat the Galaxy as if made of DM
only (in some cases, normalizing the circular velocity obtained in the model for the DM particles
to the locally measured circular velocity). In reality, the stellar and gas components dominate the potential 
well in the inner Galaxy and out to, at least, our Galactocentric distance, with, e.g., the DM contribution to
local surface mass density standing essentially within the error bar from star population counts~\cite{KG91}.

Taking an observationally driven point of view, in Ref.~\cite{PR1} (hereafter Paper~I) we reconsidered 
the problem of constructing mass models for the Milky Way, and showed that, when considering
an updated set of dynamical tracers and an efficient way of scanning a multidimensional
parameter space, a high precision determination of the local DM halo density can be obtained,
namely around 0.39~GeV~cm$^{-3}$ with a 1-$\sigma$ error bar of about 7\%. This result 
relies partially on the assumption of spherical symmetry we imposed on the DM component 
(and on the same prescription to compute the gravitational potential described in section \ref{df} of the present paper),
and would be revised if one argues, for instance, for some level of flattening of the DM halo, see, e.g.,
\cite{Salucci:2010qr,Pato:2010yq,Garbari:2011dh,Iocco:2011jz}. 
At the same time, it is a rather solid result since there is no tension 
between the simplified picture we proposed and any of the very variegated sets of complementary 
observables implemented in the analysis (this, together with the effectiveness of Bayesian inference applied to our 
mass model for the Galaxy, is reflected in the tiny error we found).

In case one assumes that the DM distribution is also isotropic, there is a 
one-to-one correspondence between the spherically symmetric density profile and the underlying 
distribution function, with the latter that can be computed from the former through Eddington's formula~\cite{BT}.
Although this requires some heavy numerical integrals, it is nowadays possible to perform this inversion
on very large samples of trial cases. Relying on the same mass models introduced in Paper~I and 
an analogous Bayesian approach with a Markov Chain Monte Carlo scan of the underlying parameter space,
we study here the family of isotropic distribution functions one can associate to the spherical DM
density profiles. Following this approach, we can address here for the first time the theoretical uncertainty 
on the direct detection signal within a framework in which the value of the local halo density, the shape of 
the distribution function and its truncation at the escape velocity, and the circular velocity of the Sun 
are taken self-consistently and in agreement with the available dynamical constraints. 
Such method is way more powerful than deriving an overall theoretical error assuming that the local halo density, 
the velocity dispersion, the escape velocity and the Sun circular velocity have given uncertainties to be propagated as 
uncorrelated errors, as actually done in most analyses in the literature. 

Results presented in this paper are valid in the limit 
of spherical symmetry\footnote{For the DM profile and for the gravitational potential, 
which has been ``symmetrized'' according to the prescription presented in section \ref{df}.} and isotropy, 
indeed rather strong assumptions, and should be regarded
as a first step towards a study allowing for, at least, axisymmetric configurations (the case with axisymmetric
models can be in principle treated in a specular way, but it is computationally much more demanding; such 
case is subject of ongoing work). On the other hand, the inner regions of a galaxy, such as at own position 
within the Milky Way DM halo, are those for which the simulations find weaker evidence for departure from
spherical symmetry, and, favored also by the presence of large amount of baryonic matter, those with the 
largest chance for gravitational relaxation of the collisionless DM system, and hence where the distribution 
function is expected to be close to isotropic. 

The paper is organized as follows: In Section~\ref{df} we introduce the Galactic model and the procedure adopted to compute the local distribution function. In Section~\ref{ddrate} we review the computation of direct detection rates,  emphasizing the connection to the distribution function, and detail the method implemented to compute the exclusion limit in one sample case, the one from the recent data release from the Xenon Collaboration. Section~\ref{mcmc} illustrates the statistical method implemented, while Section~\ref{results} contains our results. Section~\ref{concl} concludes.

\section{Galactic model and DM phase-space distribution function}
\label{df}

Assuming that the distribution of DM particles in the Galaxy is spherically symmetric and isotropic, and in the limit of spherical symmetry for 
the underlying gravitational potential for the Galaxy $\Phi(r)$, Eddington's formula~\cite{BT} gives an one-to-one correspondence between the 
DM halo density profile $\rho_h(r)$ and its phase-space distribution function $F_h$:
\begin{equation}
   F_h(\eps) = \frac{1}{\sqrt{8} \pi^2} \left[ \int_0^{\eps} \frac{d^2\rho_h}{d\Psi^2} \frac{d\Psi}{\sqrt{\eps-\Psi}} + 
   \frac{1}{\sqrt{\eps}} \left(\frac{d\rho_h}{d\Psi}\right)_{\Psi=0} \right]\,,
\label{eq:edd}
\end{equation}
where we defined the relative potential $\Psi(r) = - \Phi(r) + \Phi (r=R_{vir})$, with the virial radius $R_{vir}$ being the radius at which the 
DM halo as a virialized object is truncated. $F_h$ depends on particle velocity and position in the Galaxy only 
through the value of the relative energy $\eps = - E + \Phi (r=R_{vir}) = - E_{\rm{kin}} + \Psi(r)$, with $E$ and $E_{\rm{kin}}$, respectively, the total 
and kinetic energy. From the numerical point of view, it is actually simpler to implement Eq.~(\ref{eq:edd}) by changing the integration variable from $\Psi$ to the radius of the spherical system $r$; to compute $F_h$ at the local Galactocentric distance, it is sufficient to specify the radial dependence of $\rho_h$ and $\Phi$ from the local position in the Galaxy out to $R_{vir}$. We will then make an ansatz for a mass model for the Galaxy, 
including a parameterization for the contributions of the stellar bulge/bar and stellar disc, compare it against available dynamical constraints 
and implement a prescription to extrapolate $\Phi(r)$.

The Galactic mass model we adopt is the same introduced in Paper~I; we summarize it briefly here. The DM halo component takes the generic
form:
\begin{equation}
  \rho_h(r)=\rho^{\prime} f\left(r/a_h\right)\,,
  \label{nbody}
\end{equation}
where $f(x)$ is the function defining the shape of the DM density profile. We will consider three separate choices, namely: the profile  
originally proposed by Navarro, Frenk and White~\cite{NFW} as the universal profile describing DM halos in numerical N-body simulations,
and extensively used in the literature:
\beq
f_{NFW}(x)=\frac{1}{x(1+x)^2} \,, 
\eeq
the Einasto profile~\cite{n04,graham}, favored by the latest simulations: 
\begin{equation}
  f_{E}(x) = \exp\left[-\frac{2}{\alpha_E} \left(x^{\alpha_E}-1\right)\right]\,, 
\label{eq:einasto}
\end{equation}
and, finally, the Burkert profile~\cite{Burkert:1995yz}:
\beq
f_{B}(x)=\frac{1}{(1+x)(1+x^2)} \,,
\eeq
which has core, possibly reflecting the wash out of the central DM enhancement induced by the baryon infall.
In Eq.~(\ref{nbody}), $\rho^{\prime}$ and $a_h$ fix, respectively, a reference normalization and length scale; most often these are given 
in terms of the virial mass $M_{vir}$ and the concentration parameter $c_{vir}$, inverting the relations:
\beqra
  M_{vir} & \equiv & \frac{4\pi}{3} \Delta_{vir} \bar{\rho}_0\,R_{vir}^3 
  = \frac{\Omega_{\rm DM}+\Omega_{\rm b}}{\Omega_{\rm DM}} \,4\pi \int_0^{R_{vir}} dr \, r^2  \rho_h(r) \\
  c_{vir} & \equiv & R_{vir}/r_{-2}
\eeqra
where, in the first equation, which defines also the virial radius $R_{vir}$, the virial overdensity 
$\Delta_{vir}$ is computed according to Ref.~\cite{BN}, while $\bar{\rho}_0$ is the mean background density today.
$\Omega_{\rm DM}$ and $\Omega_{\rm b}$, the dark matter and baryon energy densities in units of the critical density,
enter in this expression since we are assuming that only a fraction equal to ${\Omega_{\rm DM}}/({\Omega_{\rm DM}+\Omega_{\rm b}})$ 
of the total virial mass consists of dark matter; their values are fixed according to the mean values from the fit of the 7-year WMAP data~\cite{WMAP7}.
Finally, in the equation for $c_{vir}$, $r_{-2}$ is the radius at which the effective logarithmic slope of the profile is $-2$.

For what regards the luminous mass components, the stellar disk mass density profile is assumed to take the form:
\beq
\rho_d(R,z) = \frac{\Sigma_{d}}{2 z_{d}} \, e^{-\frac{R}{R_d}} \, \textrm{sech}^2\left( \frac{z}{z_d}\right)
\;\;\;\; {\rm{with}} \;\;\;\; R<R_{dm}\;,
\label{disk}
\eeq 
where $\Sigma_{d}$ is the central disk surface density, $R_d$ and $z_d$ are the length scales in the cylindrical coordinates R and z, while $R_{dm}$ is the truncation radius. Among these, only $\Sigma_{d}$ and $R_d$ will be treated as free parameters. $R_{dm}$ will be assumed to scale with the local Galactocentric distance $R_0$ according to the prescription $R_{dm}= 12 [1+0.07(R_0-8~{\rm kpc})]~{\rm kpc}$~\cite{Freud}; we will also fix the vertical height scale to the best fit value suggested in Ref.~\cite{Freud},  $z_{d}$~=~0.340~kpc, since the dynamical constraints we consider in our analysis are insensitive to a slight variation around this value. The bulge/bar region is modeled through the mass density profile:
\beq
\rho_{bb}(x,y,z)= \bar{\rho}_{bb}\left[ s_a^{-1.85} \,\exp(-s_a) + \exp\left(-\frac{s_b^2}{2}\right) \right] \,
\label{bb}
\eeq
where 
\beq
s_a^2 =  \frac{q_b^2 (x^2+y^2)+z^2}{z_b^2}              
\qquad \quad  {\rm and} \quad \quad 
s_b^4 = \left[ \left(\frac{x}{x_b} \right)^2 + 
\left(\frac{y}{y_b} \right)^2\right]^2 +  \left(\frac{z}{z_b} \right)^4 \,.
\eeq
The first term in Eq.~(\ref{bb}) represents an axisymmetric nucleus~\cite{Klypin}, while the second term describes a triaxial bar. When comparing against dynamical constraints, we will implement an axisymmetrized version of Eq.~(\ref{bb}), and assume $x_b \simeq y_b = 0.9~{\rm kpc} \cdot (8~{\rm kpc}/R_0)$, $z_b=0.4~{\rm kpc} \cdot (8~{\rm kpc}/R_0)$ and $q_b= 0.6$. As for the stellar disk, we fixed these parameters because of the lack of observables to discriminate among these values and small deviations around them. Such values are in agreement with the best fit obtained in Ref.~\cite{Zhao:1995qh} assuming $R_0=8~{\rm kpc}$, and then scaled to an arbitrary $R_0$. Rather than using the two mass normalization  
scales $\Sigma_{d}$ and $ \bar{\rho}_{bb}$, in Paper~I the parameter space of our reference Galactic model was scanned varying the disc mass $M_d$ and the balge/bar mass $M_{bb}$; here we re-parameterize these in terms of two dimensionless quantities, namely, the fraction of collapsed baryons $f_{\rm b}$ and the ratio between the bulge/bar and disk masses $\Gamma$, i.e., respectively:
\beqra
  f_{\rm b} &  \equiv & \frac{\Omega_{\rm DM}+\Omega_{\rm b}}{\Omega_{\rm b}} \frac{M_{bb}+M_d+M_{\textrm{H}_\textrm{I}}+M_{\textrm{H}_2}}{M_{vir}} \label{eq:barpar1} \\
  \Gamma & \equiv & \frac{M_{bb}}{M_d}\,.
  \label{eq:barpar2}
\eeqra
In Eq.~(\ref{eq:barpar1}) we also included the sub-leading contributions to the total virial mass associated with the atomic ($\textrm{H}_\textrm{I}$) and the molecular ($\textrm{H}_2$) Galactic gas layers, with profiles as given in \cite{dame}, see also Paper~I. Finally, in our mass model, the baryons which do not collapse in the disc are distributed with the same profile of the DM component. 

The mass density profiles can be used as source terms in a Poisson equation to compute the Galactic gravitational potential. A rigorous procedure would require the solution of partial differential equations in cylindrical coordinates, a method which would be numerically challenging to apply to an extensive scan of our parameter space; moreover, it would actually give more information that those required in Eq.~(\ref{eq:edd}) which assumes a spherically symmetric $\Phi$. 
In this respect, since the goal is to use Eddington's formula at our position in the Galaxy, i.e. in the outskirts
of the stellar Galactic plane, it is a fairly good approximation to infer the local deepness of the gravitational potential well, and profile of $\Phi$ at larger radii, by
applying the following prescription\footnote{A non trivial check of the reliability of the method used in this work consists in comparing the velocity distribution in the direction perpendicular to the disk, $f_z$, obtained: 1) Within the approximation used in the paper. 2) {\it Exactly}, computing numerically the axisymmetric gravitational potential associated with our mass model for the Galaxy and then performing an Abel inversion to determine $f_z$. We made this comparison for a few representative points in our parameter scan and found only a very mild increase in the velocity dispersion, and hence a mild impact on the dark matter detection rate. 
}: First, varying the Galactocentric distance $\bar{R}$, we calculate the total mass profile $M(\bar{R})$ obtained integrating 
the sum of all density components we introduced within a sphere of radius $\bar{R}$ centered in the Galactic center. This quantity is then used in a Poisson equation in the spherically symmetric limit, whose solution $\Phi(r)$ reads:
\beq
\Phi(r) = \textrm{G}_N\,\left[ \int_r^{R_{vir}}d\bar{R}\,\frac{M(\bar{R})}{\bar{R}^2} - \frac{M(R_{vir})}{R_{vir}} \right] \,.  
\eeq

With this last step, one has all the ingredients to estimate the DM phase-space distribution function according to Eq.~(\ref{eq:edd}). In particular, since
we wish to discuss DM direct detection, we will focus on the local DM velocity distribution:
\beq
f_0(\bv) = \frac{1}{\rho_0}\,F_h(\eps_0(\bv))\,,
\eeq
having factorized out the local DM density $\rho_0 =  \rho_h(R_0) =   \int d^3v F_h(\eps_0(\bv))$, and defined 
$\eps_0(\bv)=-\frac{1}{2}|\bv|^2 + \Psi(R_0)$. One quantity which will be useful to compute to compare with previous analyses is the local velocity 
dispersion:
\beq
   \sigma_v^2(R_0) = \int d^3v \,|\bv|^2 f_0(\bv)\,,
   \label{sigmav}
\eeq
while one of the advantages of Eddington's approach is that the local escape velocity $\vesc$ is not a free parameter, being instead directly related to the 
gravitation potential through the relation:
\beq
\vesc(R_0) = \sqrt{2\Psi(R_0)} \,.
\label{vesc}
\eeq

As already mentioned, the DM velocity profile most often used in DM direct detection studies is the Maxwell-Boltzmann distribution. This is obtained from the distribution function of the isothermal sphere, taking a form in which radial and velocity dependence can be factorized:
\begin{equation} \label{eqn:Maxwellian}
   F_{is}(\eps) \propto \exp \left( \frac{3\,\eps}{\sigma_v^2}\right) \propto \rho_{is}(r) \,  \exp \left( -\frac{3\,|\bv|^2}{2\,\sigma_v^2}\right)  
   \quad \quad \textrm{for} \quad \quad |\bv| < \vesc (r)
\end{equation}
and 0 otherwise. Neglecting the small correction due to the truncation at the escape velocity (which is assumed here as an external input), this distribution has a constant velocity dispersion, which, after properly normalizing, is found
to be equal to $\sigma_v$.  The isothermal sphere profile has an asymptotically flat rotation curve, with the circular velocity at large radii tending to the value $\Theta_{\infty}=\sqrt{2/3} \, \sigma_v$. Assuming a flat rotation curve for the Milky Way down to our position in the Galaxy, that is $\Theta_{0}= \Theta_{\infty}$, one usually constrains the Maxwell-Boltzmann velocity distribution imposing: 
\beq
  \sigma_v = \sqrt{3/2}\,\Theta_0 \,.
  \label{nc}
\eeq
Taking into account the escape velocity in the normalization factor, the local DM velocity distribution in the isothermal sphere case reads:
\begin{equation}
   f_{0, is}(\bv) = \frac{1}{\Nesc} \left( \frac{3}{2 \pi \sigma_v^2} \right)^{3/2}
       \, e^{-3\bv^2\!/2\sigma_v^2} \quad \quad \textrm{for} \quad \quad |\bv| < \vesc(R_0)\,, 
\end{equation}
where $\Nesc$ is given by:
\begin{equation} \label{eqn:Nesc}
  \Nesc = \erf(\bar{z}) - 2 \bar{z} \exp(-\bar{z}^2) / \pi^{1/2}\,, 
  \quad \quad \textrm{with} \quad \quad 
  \bar{z} \equiv  \sqrt{3/2} \, \vesc /  \sigma_v\,.
\end{equation}

\section{Direct-detection rates and phase-space densities}
\label{ddrate}

The differential direct detection rate for a DM particle $\chi$ in
a given material (per unit detector mass) is given by the convolution
of the incident flux of the $\chi$ particles and the differential cross section for their
scattering off the target nucleus  $d\sigma/dQ$~\cite{jkg,dirdet}:
\begin{equation}
  \frac{dR}{dQ} = \frac{\rho_\chi}{M_N\,M_{\chi}} 
  \int_{|\bu| > \vmin} d^3u\; |\bu|\,f_\chi(\bu,t) \, \frac{d\sigma}{dQ} \,,
\label{eq:ddrate}
\end{equation}
where $Q$ is the energy deposited in the detector, $M_N$ the nucleus mass and
$\rho_\chi \cdot f_\chi$ the $\chi$ phase-space density at the detector and in its reference frame. 
The lower limit in the integral is the minimum velocity required for a $\chi$ particle to deposit the energy $Q$. Its expression depends from the kinematics of the scattering: In the case of an elastic scattering, such as 
for WIMP-nucleus interactions, the minimum velocity is given by
\beq
\vmin = \sqrt{\frac{Q M_N}{2 M_r^2}}\,,
\label{vmin-el} 
\eeq
with  $M_r = M_\chi M_N/(M_\chi + M_N)$ the $\chi$-nucleus reduced mass, being $M_{\chi}$ the mass of the particle $\chi$. If the scattering is instead inelastic, such as for inelastic DM, the minimum velocity also depends from the mass split $\delta$ between the DM particle and the final state; in this case one finds
\beq
\vmin = \frac{1}{\sqrt{2 Q M_N }}\,\left(\frac{Q M_N}{M_r} + \delta \right) \,.
\label{vmin-in} 
\eeq

In this paper we will mainly focus on spin-independent WIMP interactions. Rewriting accordingly the differential scattering cross section in terms of the WIMP-nucleon scattering cross section at zero momentum 
transfer $\sigma_{\chi n}$ and the nucleus form factor ${\mathcal F}_N^2$
\begin{equation}
\frac{d\sigma}{dQ} = \sigma_{\chi n} \frac{M_N}{2\,M_n^2\,|\bu|^2} A^2 {\mathcal F}_N^2(Q)\,,
\end{equation}
where $M_n = M_\chi m_p/(M_\chi + m_p)$ is the WIMP-nucleon reduced mass and $m_p$ the proton mass, the differential rate takes the form:
\begin{equation}
  \frac{dR}{dQ} = \frac{\sigma_{\chi n}}{2\,M_{\chi}M_n^2} A^2 {\mathcal F}_N^2(Q)
  \int_{|\bu| > \vmin} d^3u\; \frac{\rho_\chi  \cdot f_\chi(\bu,t)}{|\bu|} \,,
\label{eq:ddrate2}
\end{equation}
where A is the mass number of the target nucleus. In the prefactor there are quantities depending on the particle physics model, while the integrand depends on astrophysical quantities only, though the limit of integration contains again a dependence on $M_N$ and $M_\chi$.

The local $\chi$ phase-space density is computed assuming that the $\chi$ particles 
account for the entire DM component and is directly associated to $F_h(\eps)$:
\begin{equation} 
  \rho_\chi  \cdot f_\chi(\bu,t) = F_h\left[\eps_0(\bv)\right] \,,
  \label{eq:rho-f}
\end{equation}
with $\bv(t)=\bvobs(t) + \bu$. In the last equation $\bvobs$
takes into account the motion of the detector frame compared to the Galactic frame in which $F_h$
is computed. One needs to take into account the motion of the Local Standard of Rest (LSR),
namely the rotation of the sun around the Galactic center $\bv_{\mathrm{LSR}}$, the motion of the 
Earth around the Sun $\bV_\oplus$ and the Sun's peculiar velocity $\bv_{\odot,\mathrm{pec}}$:
\begin{equation}
  \bvobs(t)= \bv_{\mathrm{LSR}} + \bv_{\odot,\mathrm{pec}} + \bV_\oplus(t)
  \label{eq:coo1}
\end{equation}
with, in Galactic coordinates (where $\hat{\mathbf{x}}$ is the direction to
the Galactic Center, $\hat{\mathbf{y}}$ the direction of disk rotation,
and $\hat{\mathbf{z}}$ the North Galactic Pole) and using (partially) the notation of 
Ref.~\cite{Savage:2009mk}:
\beqra
&&\bv_{\mathrm{LSR}} = (0,\Theta_0,0) \,,\nonumber\\
&&\bv_{\odot,\mathrm{pec}} = (U_\odot,V_\odot,W_\odot) \simeq (10,5.2,7.2)\; {\rm km/s} \,. 
  \label{eq:coo2}
\eeqra
In the computations the motion of the Earth around the Sun has been modeled as in the \ds\ package~\cite{ds}. Neglecting here for simplicity the ellipticity of the Earth's orbit, one finds \cite{Savage:2006qr}
\beq
\bV_\oplus(t) =  V_\oplus \left[\eone \cos{\omega(t-t_1)} + \etwo \sin{\omega(t-t_1)}\right] \, ,
\label{eq:vearth}
\eeq
where $V_\oplus=29.8$ km/s and $t_1=0.218$ is the fraction of the year before the Spring equinox, while $\eone$ and $\etwo$ are the directions of the Earth's velocity at times $t_1$ and $t_1+0.25$ years.

From the previous equations, one can see that astrophysical uncertainties on the event rate computation are all encoded in one function, which we will denote by $g_{\chi}(\bu,t)$ and it is defined as follows 
\begin{equation}
  g_{\chi}(\bu,t) \equiv |\bu| \int d\Omega F_h\left[\eps_0(\bv)\right] \,.
  \label{gfun}
\end{equation}
This function is the key object in our analysis. For convenience, we will refer to it as to the DM phase-space density function, though it is actually an angular integral of the true phase-space density boosted in the detector rest frame and multiplied by $|\bu|$. 

\subsection{Differential rate and exclusion limit for the Xenon100 experiment}

In a real experiment Eq.~(\ref{eq:ddrate2}) has to be modified in order to account for experimental limitations related to the efficiency of the detector and its finite energy resolution. In discussing the impact of astrophysical uncertainties on the differential event rate, we will consider a sample toy-model case, focussing on specifications which would apply, as a first approximation, to the case of the Xenon detector, taking a constant efficiency $\epsilon=0.3$ and including finite energy resolution effects by assuming a Gaussian probability density $\xi(E,Q)$ that an event depositing an amount of energy $Q$ in the detector is detected with an associated energy in the interval $[E,\,E+dE]$ (the same assumptions are implemented for the Xenon experiment, e.g. see \cite{edsjo} and references therein). When plotting below observed differential event rates in an actual experiment, we will refer hence to the quantity:\footnote{Since we will not be discussing annual or daily modulation effects, we focus on time averaged differential event rates.}
\beq
\frac{dR}{dE} = \epsilon \int_0^\infty dQ \,\xi(E,Q)\,\frac{dR}{dQ}\left(Q\right) \,,
\label{eq:ddrate3}
\eeq  
where 
\beq
\xi(E,Q) = \frac{1}{\sqrt{2\pi\sigma_{\textrm{Xe}}^2(Q)}} e^{-(E-Q)^2/2\sigma_{\textrm{Xe}}^2(Q)}\,,
\eeq
with the energy dependent dispersion estimated for the Xenon detector:
\beq
\sigma_{\textrm{Xe}}(Q) = \sqrt{Q/\textrm{KeV}} (0.579 \,\textrm{KeV}) + 0.021 Q \,. 
\eeq

We will also discuss how astrophysical uncertainties affect the estimation of the exclusion limits in the scattering cross section -- particle mass plane for a direct detection experiment, focusing, as for the differential event rate, on the sample case applicable to the Xenon detector. To do that, however, we need to treat finite energy resolution effects in a more refined way than what just sketched above. We will follow the approach discussed in \cite{proflike}, and take into account that the quantity measured by the Xenon experiment is the number of photoelectrons associated with an event, and not the corresponding recoil energy. To translate such a signal in to a recoil energy, one has to use the conversion formula~\cite{Leff}
\beq
\nu(Q) = 3.6 \times Q\,\mathcal{L}_{eff}\,,
\eeq
which gives the number of expected photoelectrons (PE) as a function of the recoil energy. There has been some discussion in the literature on which is the correct form to assume for the relative scintillation efficiency of the detector $\mathcal{L}_{eff}$. In our analysis we will refer to its most recent determination~\cite{Leff} providing an estimate for such a quantity at energies as low as 3 KeV, while at lower energies we follow \cite{Aprile:2011hi} assuming that $\mathcal{L}_{eff}$ goes logarithmically to zero. The differential event rate is then given by the following expression
\beq
\frac{dR}{dS_1} = \epsilon\,\sum_{n=1}^{\infty}\int_{0}^{\infty} dQ\,\left[\frac{dR}{dQ}\left(Q\right)\times\frac{\nu(Q)^n}{n!}e^{-\nu(Q)}\right]\times \frac{1}{\sqrt{2\pi\sigma^2_{\textrm{PMT}}}} e^{-\frac{(n-S_1)^2}{2 n \sigma^2_{\textrm{PMT}}}}\,, 
\label{eq:ddrate4}
\eeq 
where the convolution with the Poisson distribution of mean $\nu(Q)$ accounts for the fact that a given recoil energy has a probability $\nu(Q)^n/n!\,e^{-\nu(Q)}$ of generating $n$ observable PE. The Gaussian weighted sum, instead, describes the finite resolution in the observed number of PE: the probability of observing $S_1$ PE when the actual number is $n$ is assumed to be Gaussian, with dispersion equal to $\sqrt{n}\,\sigma_{\textrm{PMT}}$ and $\sigma_{\textrm{PMT}}=0.5$ PE. Following \cite{Aprile:2011hi}, we define as signal region the interval in the observed number of photoelectrons between $4$ PE and $30$ PE. In this region the Xenon collaboration has reported three candidate DM events associated with recoil energies of $12.1$ KeV, $30.2$ KeV and $34.6$ KeV. To extract an upper bound on the cross section $\sigma_{\chi n}$ from the three Xenon100 events, one can calculate, as a function of $\sigma_{\chi n}$, the probability - denoted by $C(\sigma_{\chi n})$ - that a random experiment produces a number of events in the signal region equal or larger than four. Then, one can exclude cross sections such that, for instance, $C(\sigma_{\chi n}) \ge 0.9$. The probability $C(\sigma_{\chi n})$ can be written as the sum of different Poisson distributions of mean $\mu$: 
\beq
C(\sigma_{\chi n}) = \sum_{k=4}^{\infty} \,\frac{\mu(\sigma_{\chi n})^k}{k!}\,e^{-\mu(\sigma_{\chi n})}\,,
\eeq 
where $\mu(\sigma_{\chi n})=\mu_S(\sigma_{\chi n})+\mu_B$ is the total number of events in the signal region, including any known background term $\mu_B$. In the case of the Xenon100 experiment, we include a constant background contribution $\mu_B$ equal to 1.8 events. The signal contribution, instead, is given by:
\beq
\mu_S(\sigma_{\chi n})=\eta\,\int_{4}^{30} dS_1\,\frac{dR}{dS_1}\left(S_1\right)\,,
\eeq
where $\eta=48\times 100.9$ [kg days] is the exposure of the Xenon100 experiment. 
Introducing the function: 
\beq
W(\sigma_{\chi n})=C(\sigma_{\chi n})-0.9\,,
\eeq 
we can define the Xenon100 upper limit $\bar{\sigma}(M_{\chi})$ as:\footnote{This condition defines the frequentist 90$\%$ confidence level exclusion limit. In the context of the present Bayesian analysis, this relation has to be regarded simply as our definition of the exclusion limit as a function of the DM particle mass and Galactic model parameters.} 
\beq
W(\bar{\sigma}(M_{\chi})) = 0\,,
\label{limit}
\eeq 
where in the last line we emphasized the dependence of the exclusion limit $\bar{\sigma}$ from the DM particle mass.

\section{Bayesian analysis of the parameter space}
\label{mcmc}
Eddington's inversion formula Eq.~(\ref{eq:edd}) establishes - under the assumptions discussed - a direct relation between the local DM phase-space density and the parameter space of the underlying Galactic model. 
In turn, the local DM phase-space density, along with the coordinate transformation in Eqs.~(\ref{eq:coo1}) and~(\ref{eq:coo2}), defines the factor in the DM scattering rate depending on astrophysical assumptions.  
To address the astrophysical uncertainties on DM direct detection we need therefore to study the variability range of the Galactic parameters. The model we summarized in Section~\ref{df} has 7 or 8 free parameters:
the local Galactocentric distance $R_0$; the DM halo parameters $c_{vir}$ and $M_{vir}$, plus the exponent $\alpha_E$ in case of the Einasto profile; the parameters defining mass profiles for the baryonic components $R_d$, $f_b$ and  $\Gamma$ (recall Eqs.~(\ref{eq:barpar1}) and (\ref{eq:barpar2})); the halo star anisotropy parameter $\beta$ (introduced to make predictions for the velocity dispersion of halo stars and then compare it to the data of \cite{SDSS6}, see Paper~I for further details). To set constraints on this parameter space, we consider a broad and variegated sample of  dynamical constraints for the Milky Way, including several recent results, the same set of observables implemented in the analysis on the local halo density of Paper~I (to which we refer for a comprehensive description): the proper motion of stars in the outer Galaxy, the radial velocity dispersion of halo stars, terminal velocities for the rotation curve in the inner part of the Galaxy, the monitoring of stellar orbits around the Galactic Center, the peculiar motion of Sgr$A^*$, the Oort's constants,  the total mean surface density within $|z| < 1.1 \rm kpc$, the local disk surface mass density and, finally, the total mass inside 50 kpc and 100 kpc.

The scan of the parameter space to address variability regions of the parameters is performed implementing a Bayesian analysis, again specular to the one carried out in Paper~I. We sample the Galactic model posterior probability density function (pdf) implementing a Markov Chain Monte Carlo scan of the underlying parameter space. By construction, the pdf reflects the change of our prejudice about the most probable values for the model parameters ${\bf g}$ after seeing the data $d$. It is related to the Likelihood $\mathcal{L}(d|{\bf g})$ and the prior distribution $\pi({\bf g})$ through Bayes' theorem, which states      
\beq
p({\bf g}|d) = \frac{\mathcal{L}(d|{\bf g}) \pi({\bf g})}{p(d)} \,,
\label{Bayes}
\eeq
where p(d) is the Bayesian evidence and in the present discussion simply plays the r\'ole of a normalization constant. For a detailed description of our choice of the Likelihood and of the prior distribution we refer the reader to Paper~I. In the present work, we generated 16 chains for each DM profile, accumulating (after burn in) approximately $1.5\times10^5$ accepted points in the parameter space for every considered profile. 

Rather than to the posterior pdf of the Galactic model parameters, in this work we are interested in the marginal one or two-dimensional pdf of specific functions of the Galactic model parameters. We therefore used in our analysis the fact that the marginal posterior pdf for any quantity $f$ depending on the model parameters - denoted by $p(f,{\bf g}|d)$ - is related to the posterior pdf $p({\bf g}|d)$ by the relation
\beq
p(f,{\bf g}|d) = \delta(f({\bf g})-f)\,p({\bf g}|d) \,,
\label{fg}
\eeq
which follows from the definition of conditional pdf. Expectation values and variances of any function $f$ of the model parameters with respect to $p({\bf g}|d)$ can be then calculated as follows  
\beq
\langle f \rangle = \int d{\bf g} \,f({\bf g})\,p({\bf g}|d)\,; \qquad\quad 
\sigma_{f}^2 = \langle f^2 \rangle-\langle f \rangle^2 \,. 
\label{ev}
\eeq 

Eqs.~(\ref{fg}) and (\ref{ev}) apply to functions of the Galactic model parameters only. The phase-space density, the direct detection differential event rate and the Xenon100 exclusion limit are instead all functions of both the Galactic model parameters {\it and} a further continuous variable. Respectively: the detector rest frame velocity $u$ \footnote{Indeed, the phase-space density is also a function of time (see Eq.~(\ref{gfun})). In all figures shown in this section we consider a time average of the the function $g_{\chi}$.}, the observed recoil energy $E$ and the DM particle mass $M_{\chi}$. To keep the present discussion as general as possible, we will denote during this section such an additional continuous variable by $x$. 

The problem is now that of determining the marginal posterior pdf of objects $\zeta$ - functions of the Galactic model parameters - with an explicit dependence from $x$, {\it i.e.} $\zeta : x \longrightarrow \zeta(x)$. In our analysis we compute such pdf's discretizing the variable $x$ (see also \cite{Arina:2011si} for a different approach to this problem). We therefore use Eq.~(\ref{fg}) to calculate the marginal posterior pdf for a set of $N$ quantities $\zeta_{i} = \zeta(x_i)$, with $i=1,\cdots N$, where now each $\zeta_{i}$ is a function of the Galactic model parameters only. If the steps of such discretization $\{ x_1 \cdots x_N \}$ are small enough, this procedure leads to a good determination of the function $\zeta(x)$. As a result, to any given point $x_{k}$ in the range of variability of x, we can associate instead of a single number, a marginal posterior pdf, namely the pdf related to $\zeta_{k}$ through Eq.~(\ref{fg}).

Finally, given a marginal posterior pdf for the quantities $\zeta_i$, one can calculate from Eq.~(\ref{ev}) the means $\langle \zeta_i  \rangle$ and construct the required $\gamma\%$ credibility intervals, namely the shortest intervals including the $\gamma\%$ of the total probability. In our analysis we considered $\gamma\%$ equal to either 68$\%$ or 95$\%$. 

\section{Results}
\label{results}
The main result of the present analysis is a Bayesian determination of the local DM phase-space density based on the Eddington's inversion formula and a broad sample of dynamical constraints for the Milky Way. As already mentioned, this quantity incorporates all the astrophysical uncertainties which affect the expected event rate at a DM direct detection experiment. We therefore first focus on this quantity; then, in a few illustrative examples, we also show how astrophysical uncertainties translates in to error bars for the predicted differential event rate and exclusion limit in the scattering cross section - WIMP mass plane. Concerning the exclusion limit, we will concentrate on the last Xenon100 data release, however, our results for the phase-space density can be readily applied to any other case.

\subsection{Phase-space density}
\begin{figure}
\includegraphics[width=50mm,height=50mm]{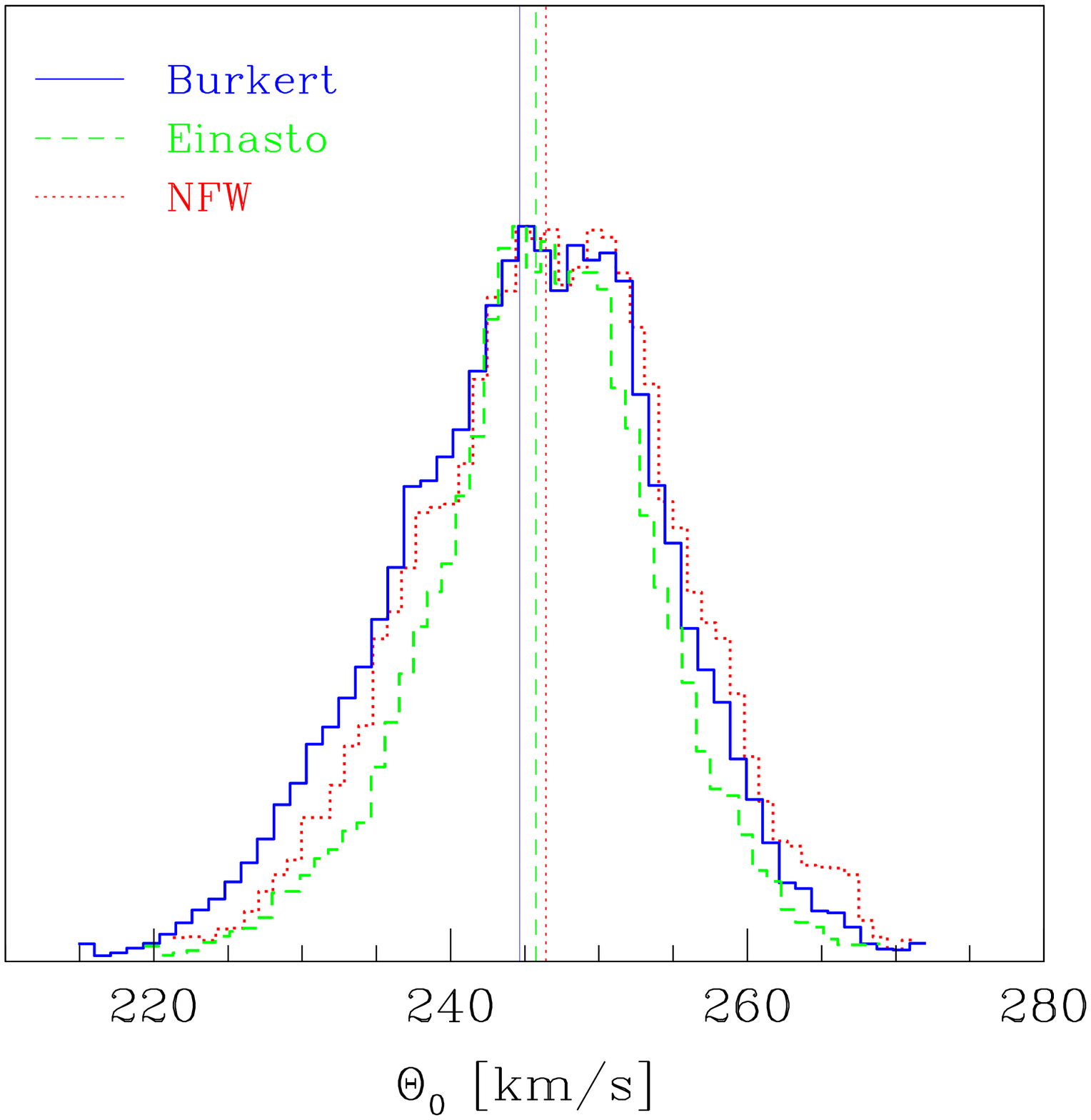}
\includegraphics[width=50mm,height=50mm]{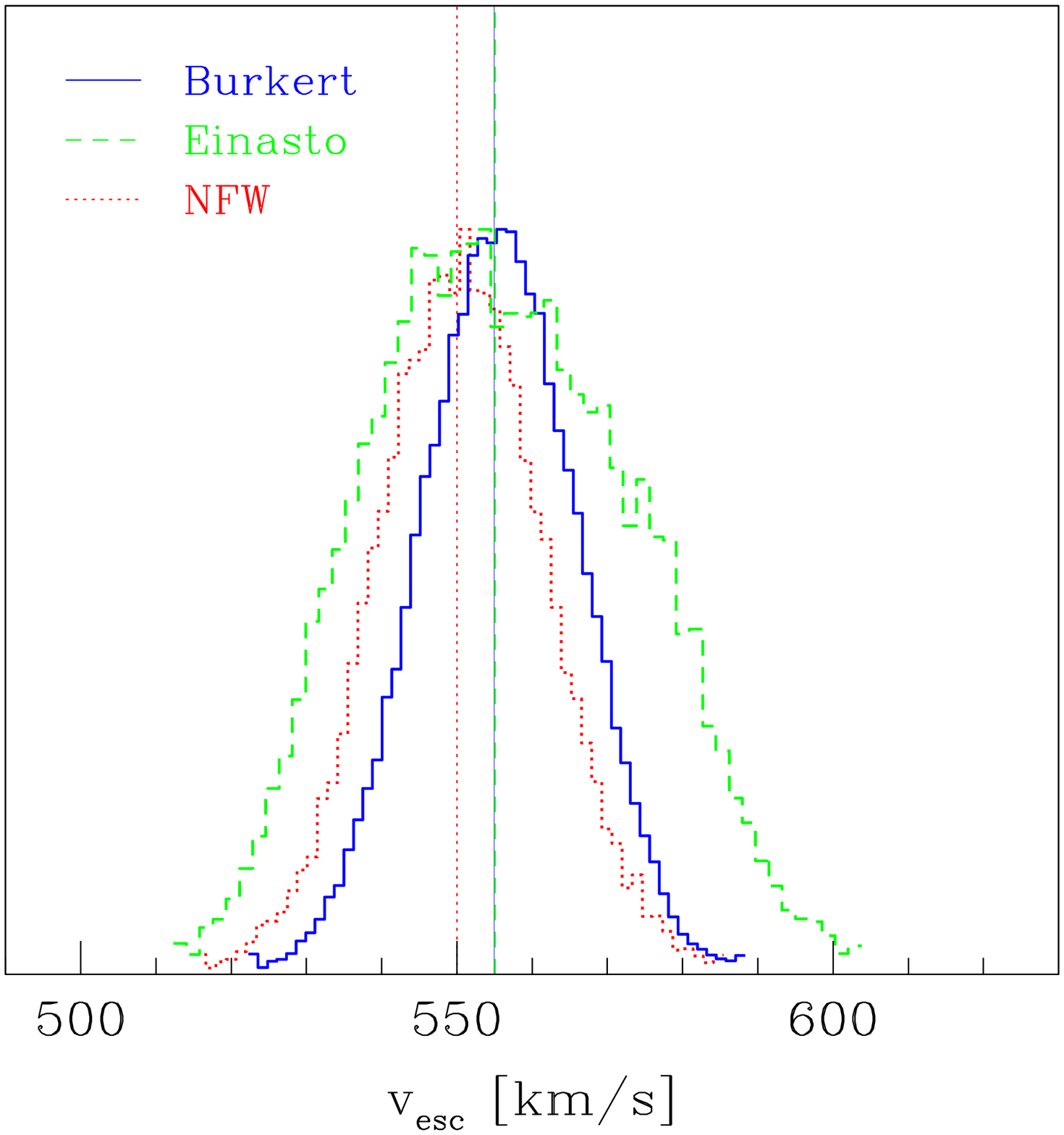}
\includegraphics[width=50mm,height=50mm]{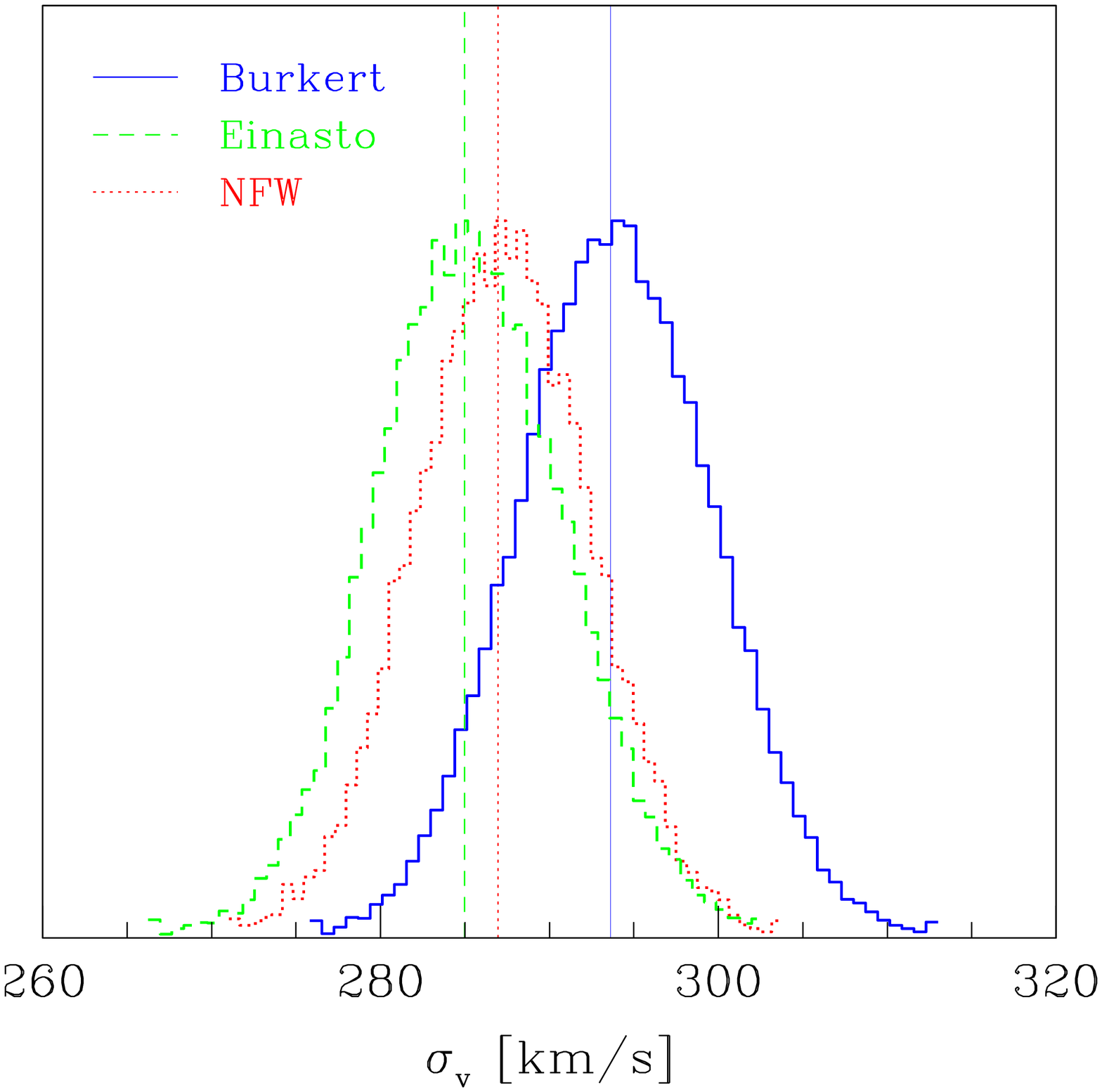}
\caption{Left panel: marginal posterior pdf for the local circular velocity. Central panel: marginal posterior pdf for the local escape velocity. Right panel: marginal posterior pdf for the local velocity dispersion.}
\label{vel123}
\end{figure}

To better understand our results for the DM phase-space density, it is instructive to analyze first a few related local quantities, namely the circular velocity $\Theta_{0}$, the escape velocity $\vesc$ and, finally, the DM velocity dispersion $\sigma_v$. 
%---------------------------Table--------------------------------------
\begin{table}[t]
  \footnotesize
  \centering
  \renewcommand{\arraystretch}{1.}
  \begin{center}
    {\bf NFW}
  \end{center} 
  \begin{tabular}{|c|c|c|c|c|c|c|}
    \hline
   Parameters               &     mean      &  $\sigma$ & lower 68\% & upper  68\% & lower 95\% & upper 95\% \\
    \hline \hline
 $f_b$              &   0.25    &  0.039   & 0.21    &  0.29    &  0.17   &  0.33 \\  
 $\Theta_{0}$ [km/s]&   246.3   &  8.7     & 237.6   &  254.8   &  228.9  &  263.6 \\
 $\vesc$ [km/s]     &   550.7   &  11.2    & 539.7   &  561.7   &  528.5  &  573.3 \\
 $\sigma_v$ [km/s]  &   287.0   &  5.2     & 281.7   &  292.2   &  276.7  &  297.2 \\
  \hline 
  \end{tabular}
  \footnotesize
  \centering
  \renewcommand{\arraystretch}{1.}
  \begin{center}
    {\bf Burkert}
  \end{center} 
  \begin{tabular}{|c|c|c|c|c|c|c|}
    \hline
   Parameters               &     mean      &  $\sigma$ & lower 68\% & upper  68\% & lower 95\% & upper 95\% \\
    \hline \hline
 $f_b$              &   0.36    &  0.049   & 0.31    &  0.41    &  0.27   &  0.46 \\  
 $\Theta_{0}$ [km/s]&   244.8   &  9.1     & 235.7   &  253.5   &  226.0  &  261.5 \\
 $\vesc$ [km/s]     &   555.4   &  10.5    & 545.0   &  566.0   &  534.9  &  575.8 \\ 
 $\sigma_v$ [km/s]  &   293.6   &  5.77    & 287.8   &  299.3   &  282.4  &  304.9  \\
  \hline 
  \end{tabular}
  \footnotesize
  \centering
  \renewcommand{\arraystretch}{1.}
  \begin{center}
    {\bf Einasto}
  \end{center} 
  \begin{tabular}{|c|c|c|c|c|c|c|}
    \hline
   Parameters               &     mean      &  $\sigma$ & lower 68\% & upper  68\% & lower 95\% & upper 95\% \\
    \hline \hline
 $f_b$              &   0.23    &  0.10  &  0.13      &  0.34    &  0.11   &  0.46 \\
 $\Theta_{0}$ [km/s]&   245.8   &  7.7   &  238.4     &  253.1   &  229.4  &  260.6 \\
 $\vesc$ [km/s]     &   555.0   & 17.6   & 536.7      &  574.0   &  523.5  &  589.2 \\
 $\sigma_v$ [km/s]  &   285.1   & 570.3  & 279.5      &  290.8   &  274.0  &  296.4 \\
  \hline 
  \end{tabular}
  \caption{Means, standard deviations and confidence intervals for the circular velocity $\Theta_{0}$, the escape velocity $\vesc$ and the dark matter velocity dispersion $\sigma_v$. We also show our results for the parameter $f_b$, which affects the shape of the local gravitational potential and therefore the local dark matter phase-space density. Other Galactic model parameters have been already discussed in Paper~I. \label{tab_all}}
\end{table}
%------------------------End Table-------------------------------------
In Table \ref{tab_all} we show means and credibility intervals for these quantities, while their marginal posterior pdf are shown in Fig.~\ref{vel123}, where each quantity has been plotted in three cases, corresponding to the three DM profiles considered in this work. As we explained in detail in the previous section, such pdf's have been determined applying Eq.~(\ref{fg}) to the case of the posterior pdf of the Galactic model parameters sampled through our MCMC scan. 

The local circular velocity (Fig.~\ref{vel123}, left panel) has been very well reconstructed in all the three cases. Indeed the three means of $\Theta_{0}$ are all close to the corresponding experimental value which has been implemented in the Likelihood function, {\it i.e.} $\Theta_0 = (245 \pm 10.4)$ km/s \cite{Reid}. The minor differences between the three cases are related to the fact that different profiles have slightly different mean values for $R_0$ and also different total mass within $R_0$. 

For what concern the escape velocity (Fig.~\ref{vel123}, central panel), we find a mean value close to 555 km/s for the Einasto and the Burkert profiles,  and to  550~km/s in the NFW case. Interestingly enough, although we did not consider any direct constraint on the escape velocity, our determination of this quantity is in a very good agreement with the central value quoted by the RAVE collaboration \cite{RAVE} (median likelihood of 544~km/s
and 90$\%$ confidence interval in the range 498--608~km/s). We also notice that in the Einasto case the $\vesc$ pdf has a broader shape. This result is related to the larger uncertainties - compared to the other cases - in the fraction of collapsed baryons that we find in a scan based on an Einasto profile (see Table \ref{tab_all}). 

The DM velocity dispersion is connected to the width of the phase-space density as a function of the velocity. From Fig.~\ref{vel123} (right panel), we can see that in the case of a Burkert profile the mean value of $\sigma_v$ is significantly larger that in the Einasto and NFW cases. This result can be understood as follows. A spherically symmetric gravitational potential generates a velocity dispersion for the local population of DM particles which can be estimated through the Jeans equation. One finds
\beq
\sigma_v^2(r) = \frac{1}{\rho_{h}(r)} \int_r^{\infty}d\tilde{r}\; \rho_{h}(\tilde{r}) \frac{d\Phi}{d\tilde{r}}\,. 
\label{jeans2}
\eeq 
By numerical inspection we find that the first derivative of the gravitational potential is larger - in the relevant integration region - for a Burkert profile than in the other cases. This is related to the fact that, according to our posterior pdf, this profile is in a better agreement with observations when a large fraction of collapsed baryons is considered, as one can see in Table \ref{tab_all}. A larger value of $f_b$ - with respect to the other profiles - is in turn responsible for the larger velocity dispersion which emerges from our marginal posterior pdf\footnote{ In this work $\sigma_v$ has been calculated through Eq.~(\ref{sigmav}). The Jeans equation is however useful to qualitatively understand the results}. 
\begin{figure}
\includegraphics[width=50mm,height=50mm]{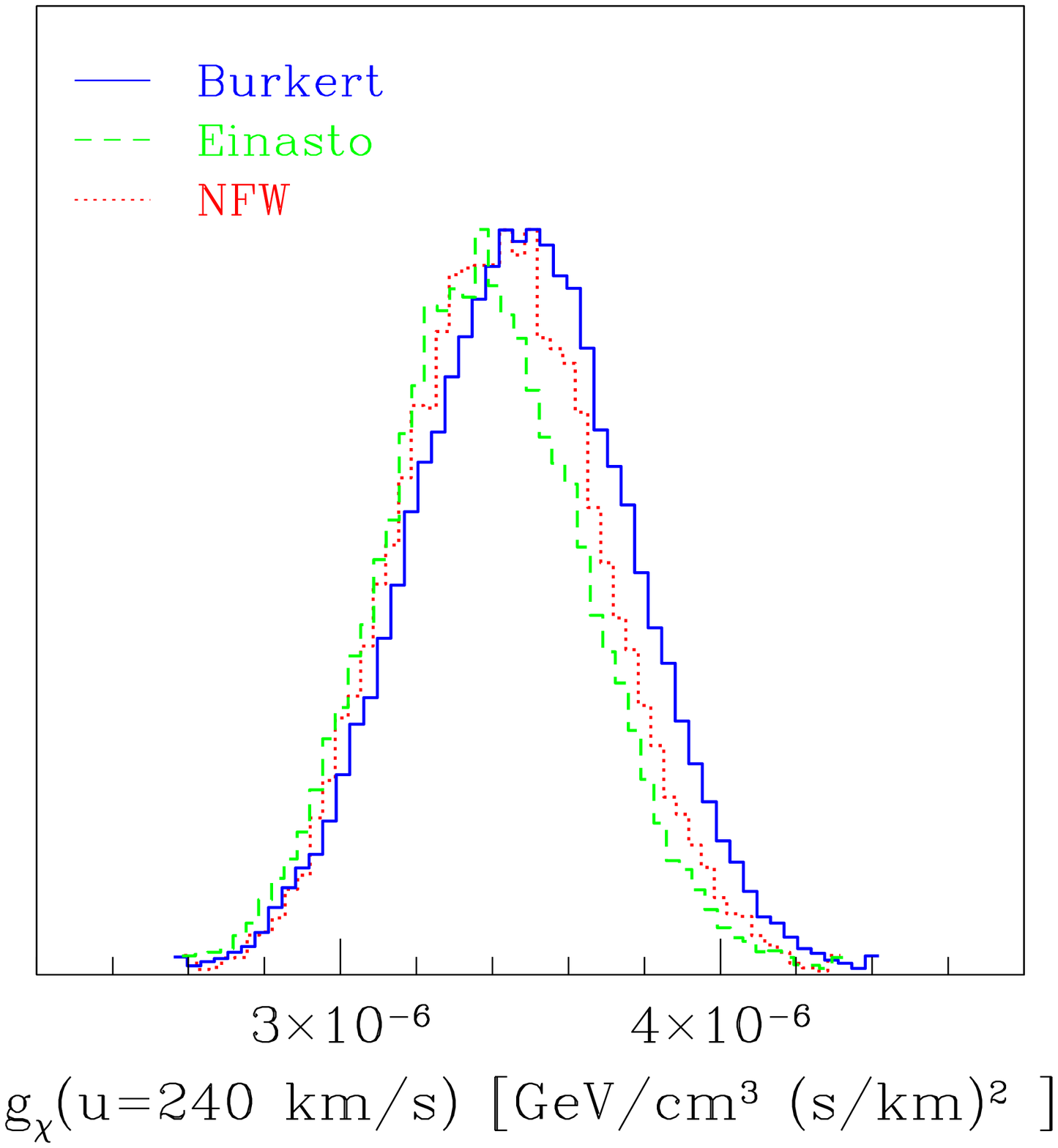}
\includegraphics[width=50mm,height=50mm]{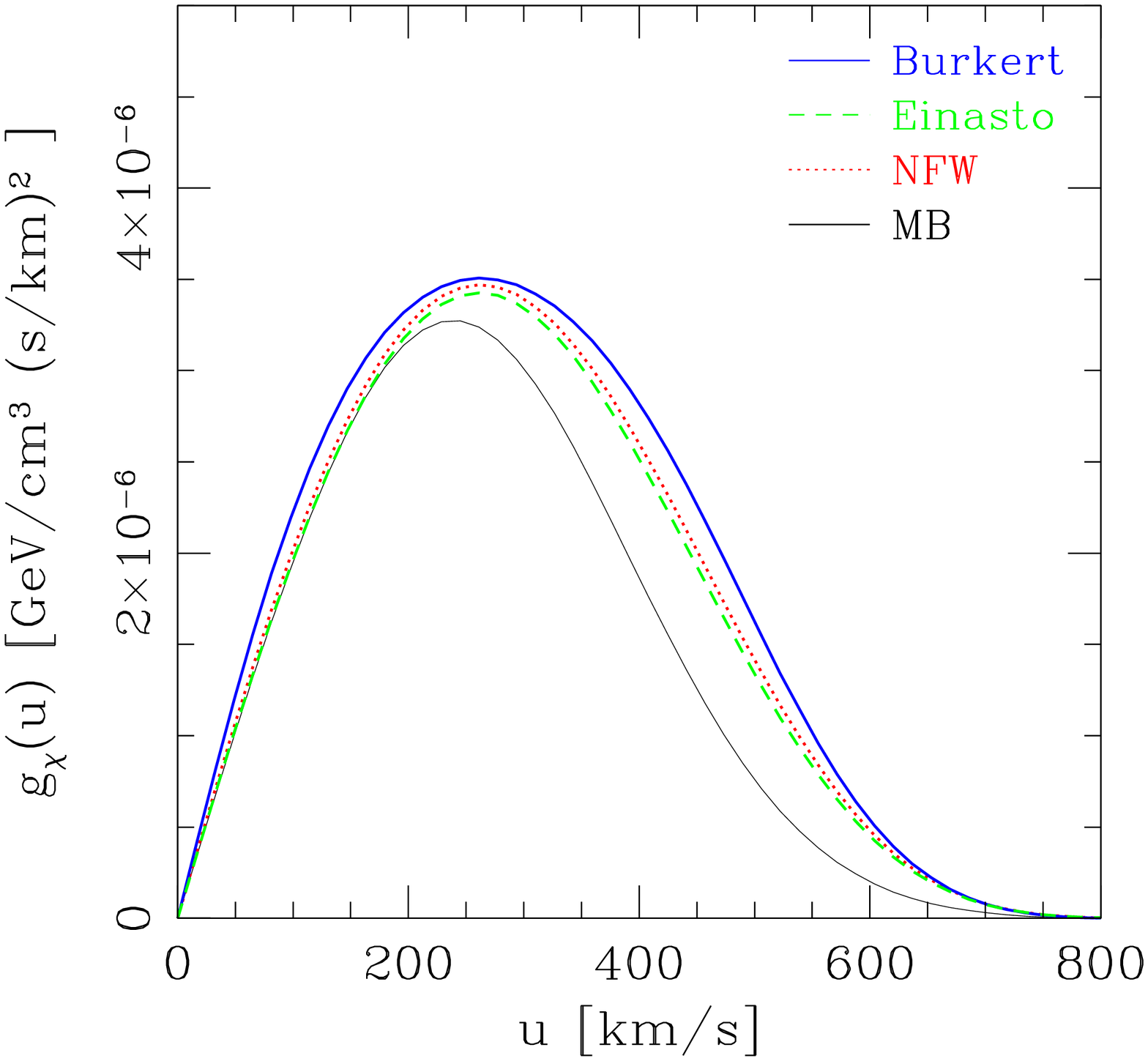}
\includegraphics[width=50mm,height=50mm]{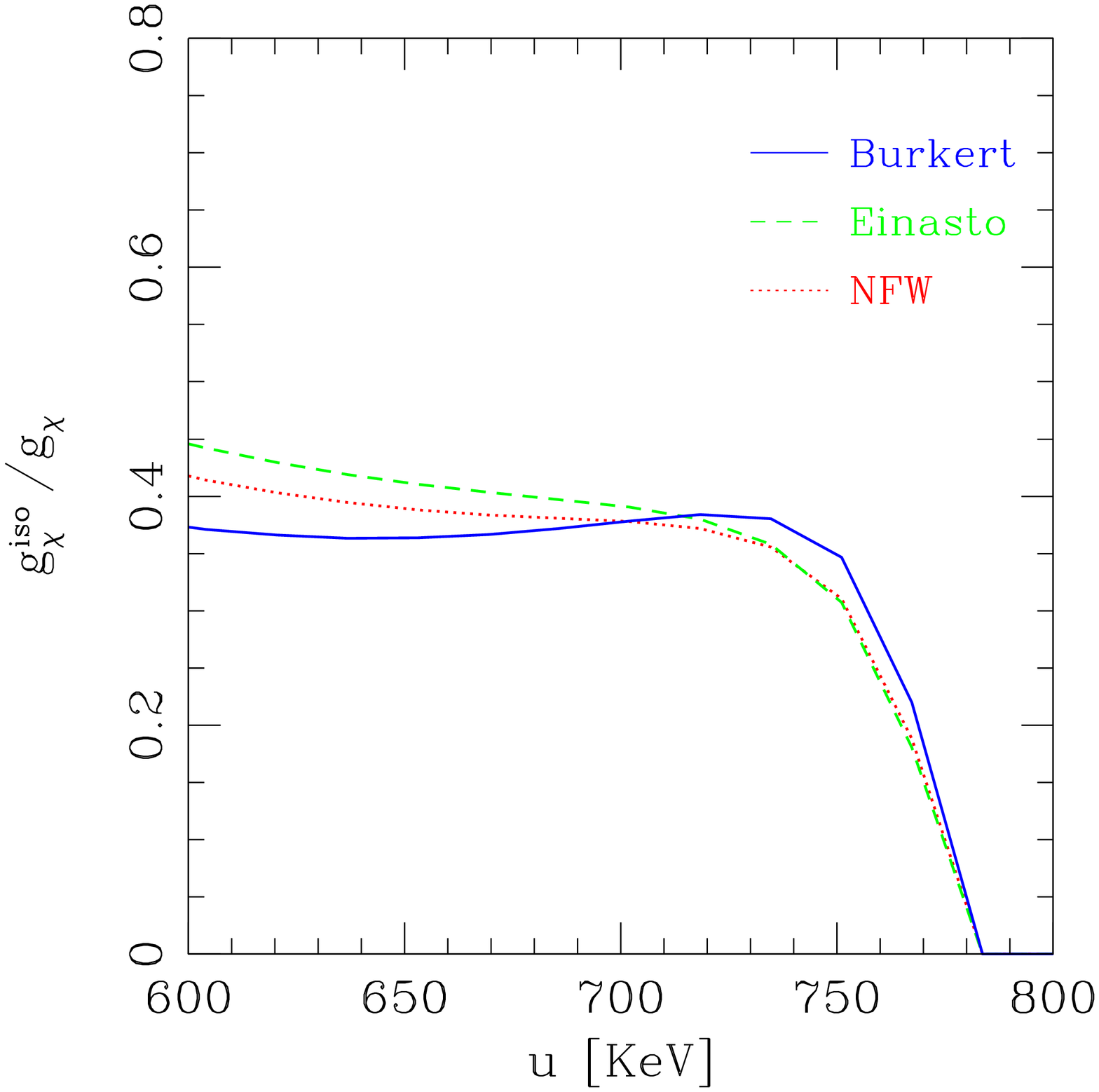}
\caption{Left panel: marginal posterior pdf for the dark matter phase-space density $g_{\chi}$ at $u=240$ km/s. Central panel: mean dark matter phase-space density. Three profiles and Maxwell-Boltzmann approximation with $\Theta_0=220$ km/s, $\vesc=544$ km/s, $\rho_{\chi}=0.3$ GeV/cm$^3$ and $\sigma_v = \sqrt{3/2}\,\Theta_0$. Right panel: ratio between the same Maxwell-Boltzmann distribution of the central panel and the tails of our MCMC distributions.}
\label{udf-mean}
\end{figure}

\begin{figure}
\includegraphics[width=50mm,height=50mm]{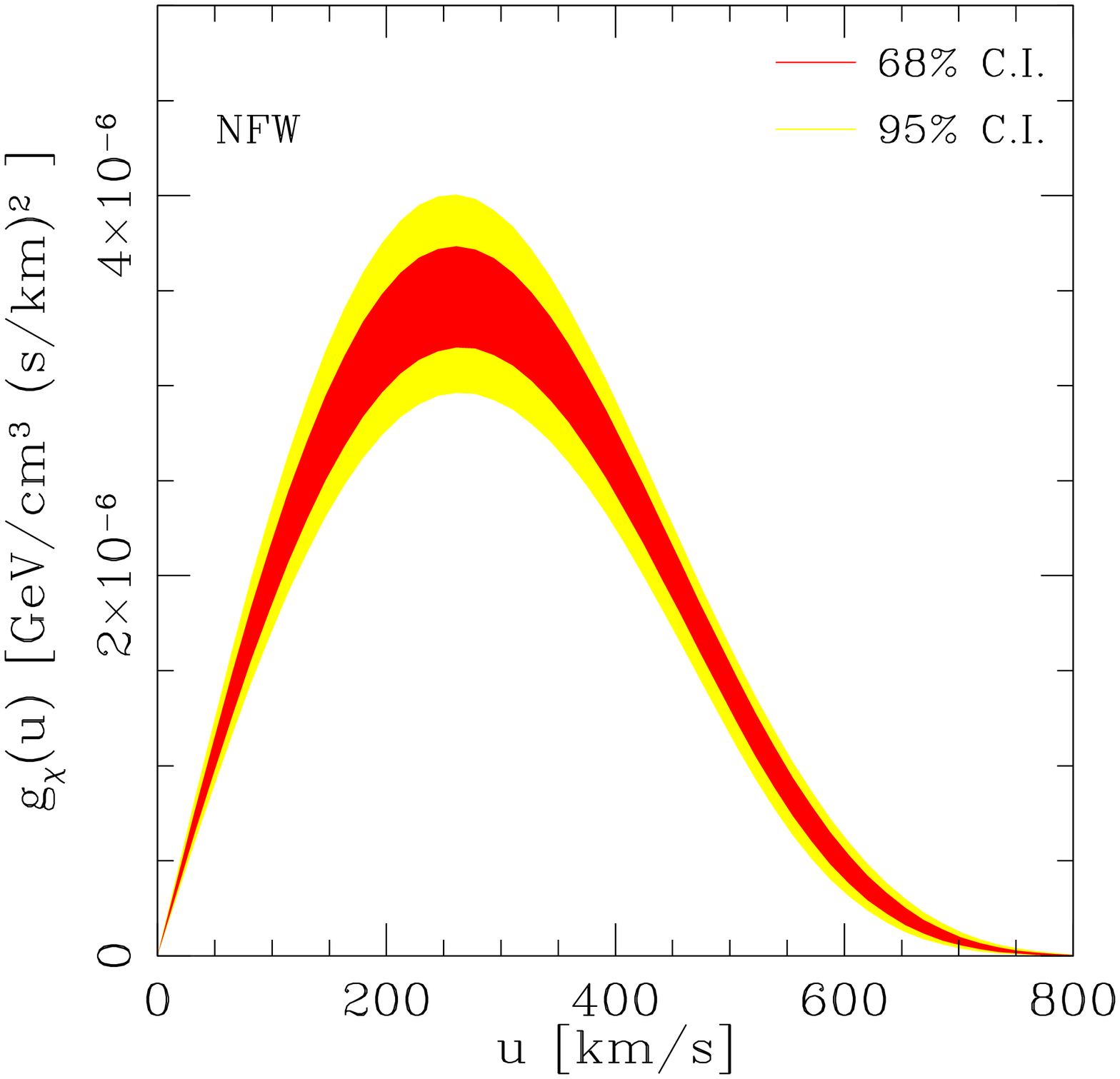}
\includegraphics[width=50mm,height=50mm]{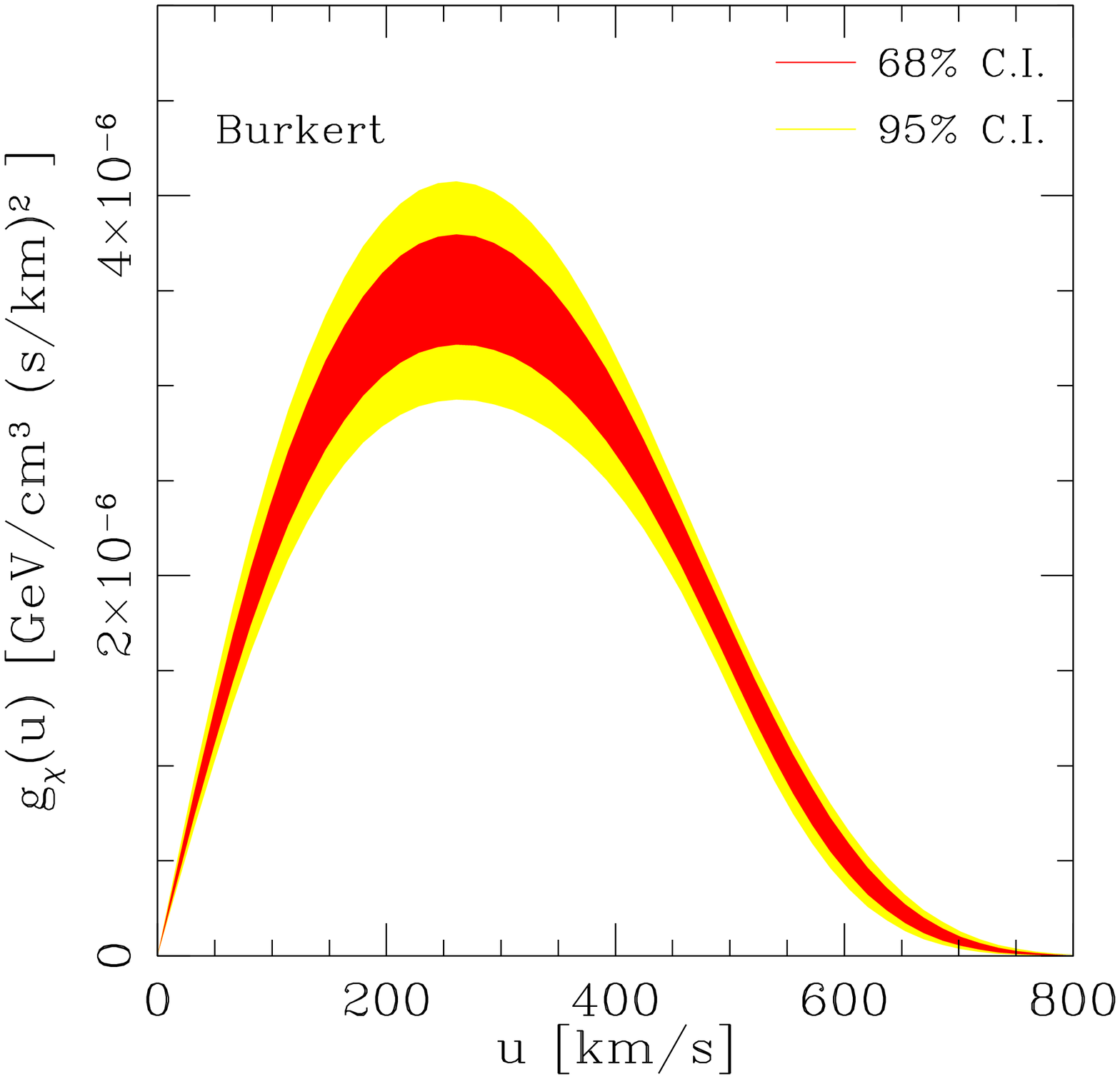}
\includegraphics[width=50mm,height=50mm]{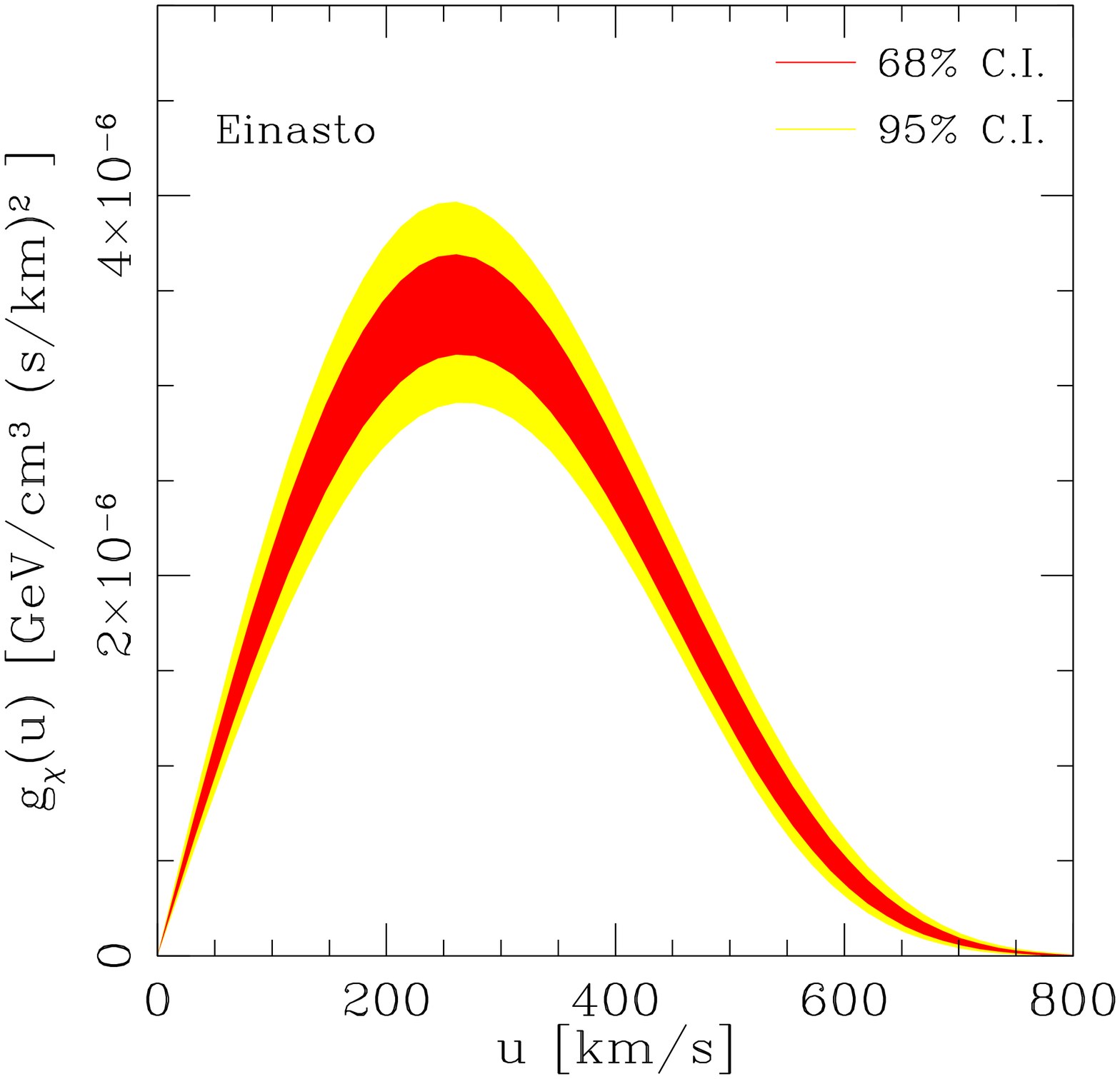}
\caption{Phase-space density ``bands'' corresponding to 68$\%$ and 95$\%$ credibility intervals. Left panel: NFW profile. Central panel: Burkert profile. Right panel: Einasto profile.}
\label{udf-all}
\end{figure}

We can now focus on the DM phase-space density. In Fig.~\ref{udf-mean} (left panel) we show the marginal posterior pdf for the DM phase-space density evaluated at a reference velocity in the detector rest frame (240 km/s). In the notation of the previous section, we are therefore considering the case in which $x=u$ and $\zeta(x)=g_{\chi}(u)$. The $u$ variable has been discretized and the reference velocity corresponds to one of the points of such a N-point discretization, {\it i.e.} $u_k=240$ km/s with $k\in\{1\cdots,N\}$. Let us note that the Burkert pdf is systematically shifted towards larger velocities. This is a feature which is present at each point of the discretization. From this marginal posterior pdf (Fig.~\ref{udf-mean}, left panel) we can determine the mean value of $g_{\chi}(u_k)$ and the corresponding 68$\%$ and 95$\%$ credibility intervals. By proceeding in this way for all the N points in the discretization we can plot a phase-space density ``band'', instead of the usual phase-space density curve. This has been done in Fig.~\ref{udf-all} for the three profiles under analysis. The red band is associated with the 68$\%$ credibility interval, while the yellow one corresponds to the 95$\%$ credibility interval. This figure makes quantitative what we already mentioned in the previous section: given a value of the velocity, say $u$, we can now associate to $u$ a pdf - and not just a number - which accounts for astrophysical uncertainties in the determination of the phase-space density at that point. 

In Fig. \ref{udf-mean} (central panel) we show the "mean" phase-space densities corresponding to the three profiles considered in this paper. In this figure to each point of the velocity discretization we associated the mean of the corresponding pdf. It should be now clear that, because of a larger velocity dispersion, the Burkert case is characterized by a broader phase-space density.

We now compare our MCMC predictions with the usual Maxwell-Boltzmann approximation. As already mentioned, a Maxwell-Boltzmann distribution is identified by two parameters: the escape velocity $\vesc$ and the velocity dispersion $\sigma_v$. Then, as explained in section \ref{df}, the Maxwell-Boltzmann distribution is normalized imposing $\sigma_v = \sqrt{3/2}\,\Theta_0$, and this parameter, together with the sharp cutoff imposed at the assumed value for the escape velocity, sets the shape of the high velocity tail of the distribution. In Fig.~\ref{udf-mean} (right-panel) we show the ratio between the distribution function found for an isothermal sphere profile (Maxwell-Boltzmann distribution times the local density, see Eq.~(\ref{eq:rho-f})) with $\Theta_0=220$ km/s, $\vesc=544$ km/s, $\rho_{\chi}=0.3$ GeV/cm$^3$ and $\sigma_v = \sqrt{3/2}\,\Theta_0$, and the tails of our MCMC phase-space densities. From this figure one can see that different profiles lead to distributions behaving differently at high velocities, crossing each other in a non trivial way. Interestingly enough, we also notice that, in the high velocity tail, our MCMC mean distributions are systematically above the ones obtained in the Maxwell-Boltzmann approximation considered in this example (and often adopted in the literature as a benchmark distribution), with a mismatch which is significantly larger than the mismatch in the numerical value for the local halo density (which is about 0.4~GeV/cm$^3$ for the three profiles considered).

\begin{figure}
\includegraphics[width=50mm,height=50mm]{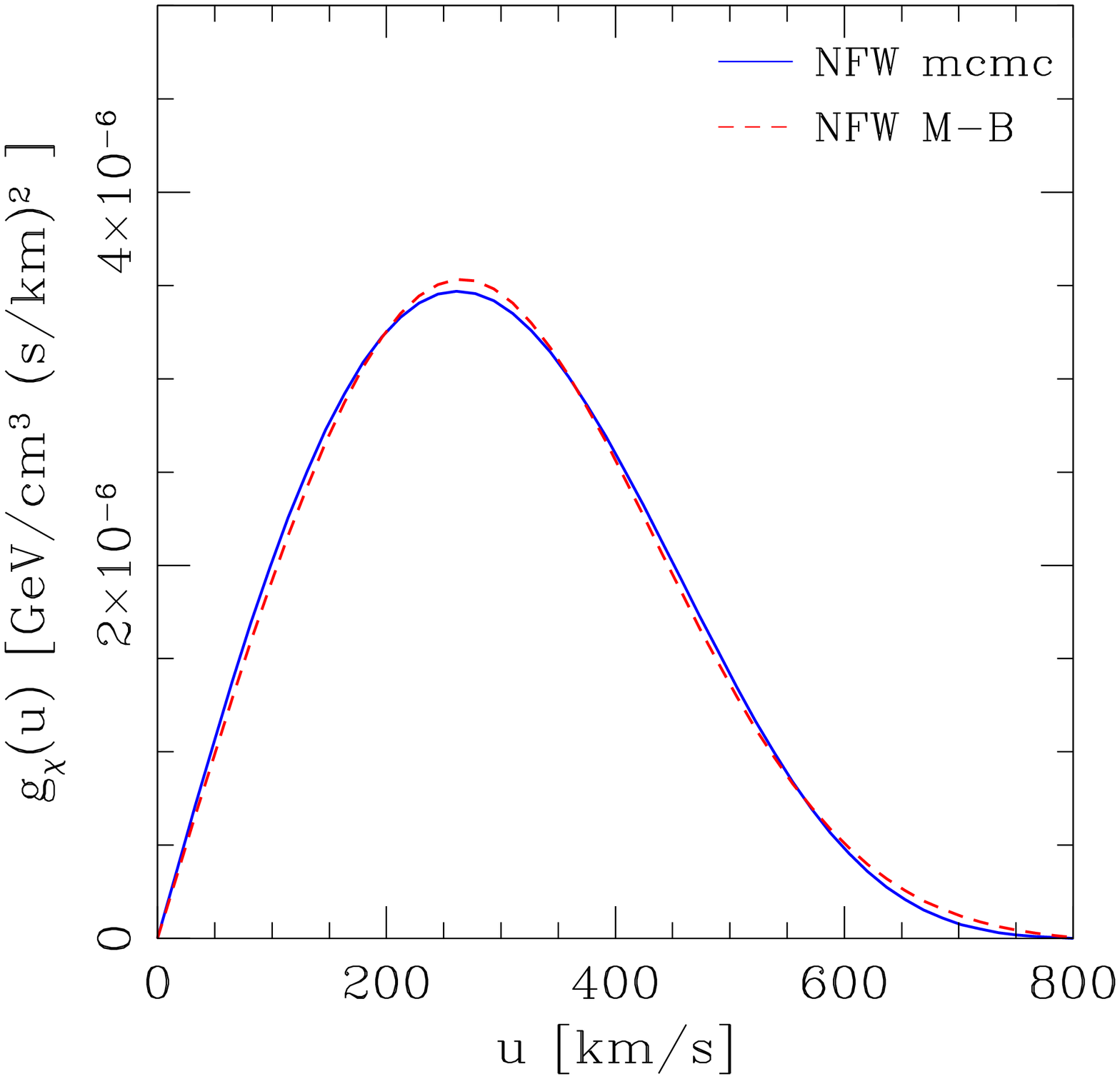}
\includegraphics[width=50mm,height=50mm]{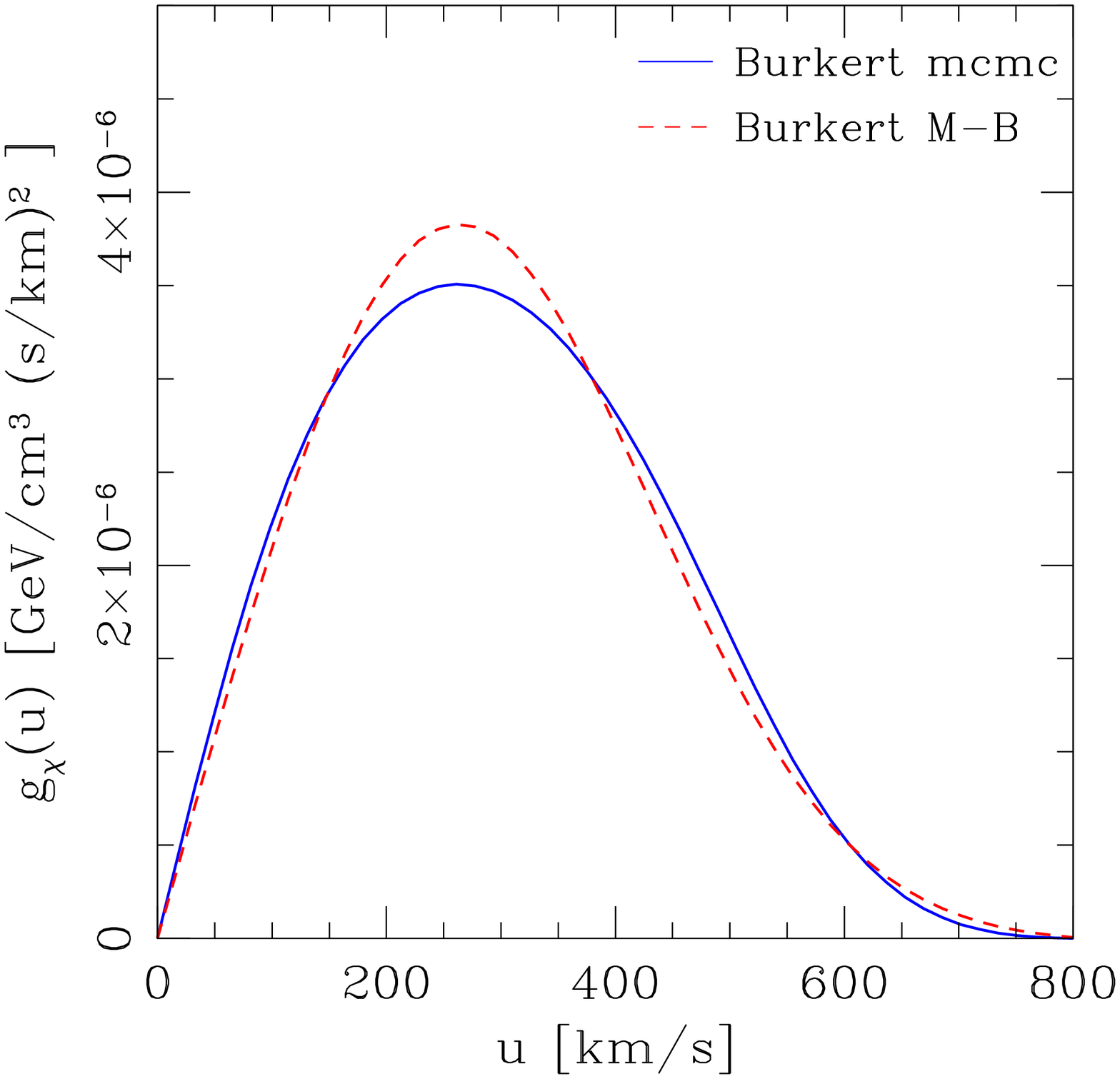}
\includegraphics[width=50mm,height=50mm]{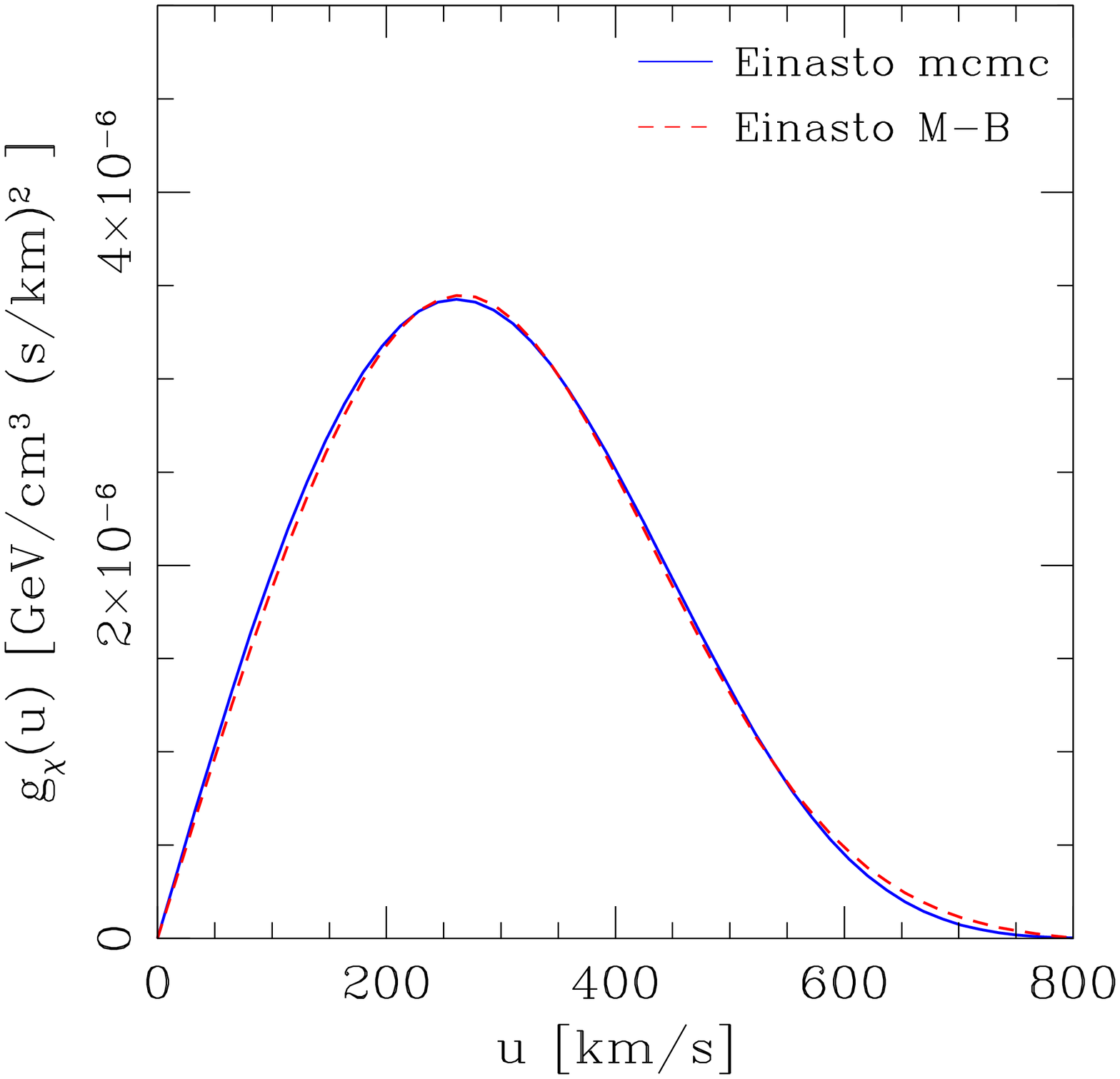}
\caption{Mean MCMC dark matter phase-space density vs. Maxwell-Boltzmann approximation (see the text for more details). Left panel: NFW profile. Central panel: Burkert profile. Right panel: Einasto profile.}
\label{MB}
\end{figure}
In Fig.~\ref{MB} we compare our mean MCMC phase-space densities with the distribution functions of different isothermal spheres, one for each profile, constructed setting the required Galactic parameters at their mean values according to our MCMC scan (see Table \ref{tab_all}). From this figure we can appreciate the fact that the phase-space density associated with a Burkert profile has a shape which - compared to the other cases - is more difficult to approximate with the distribution function of an isothermal sphere. As already pointed out, this result is related to the higher velocity dispersion found in the Burkert case. 

\begin{figure}
\includegraphics[width=50mm,height=50mm]{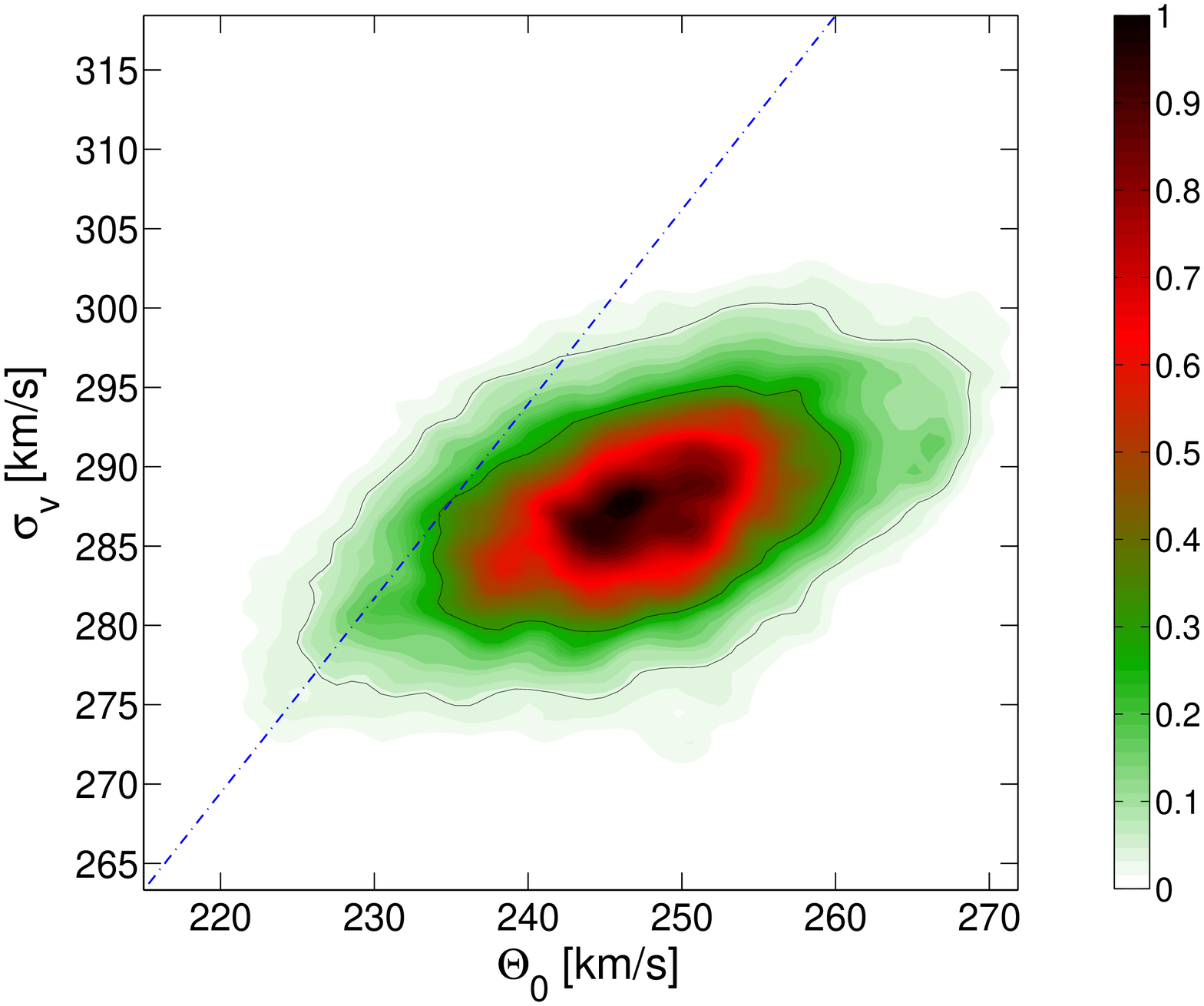}
\includegraphics[width=50mm,height=50mm]{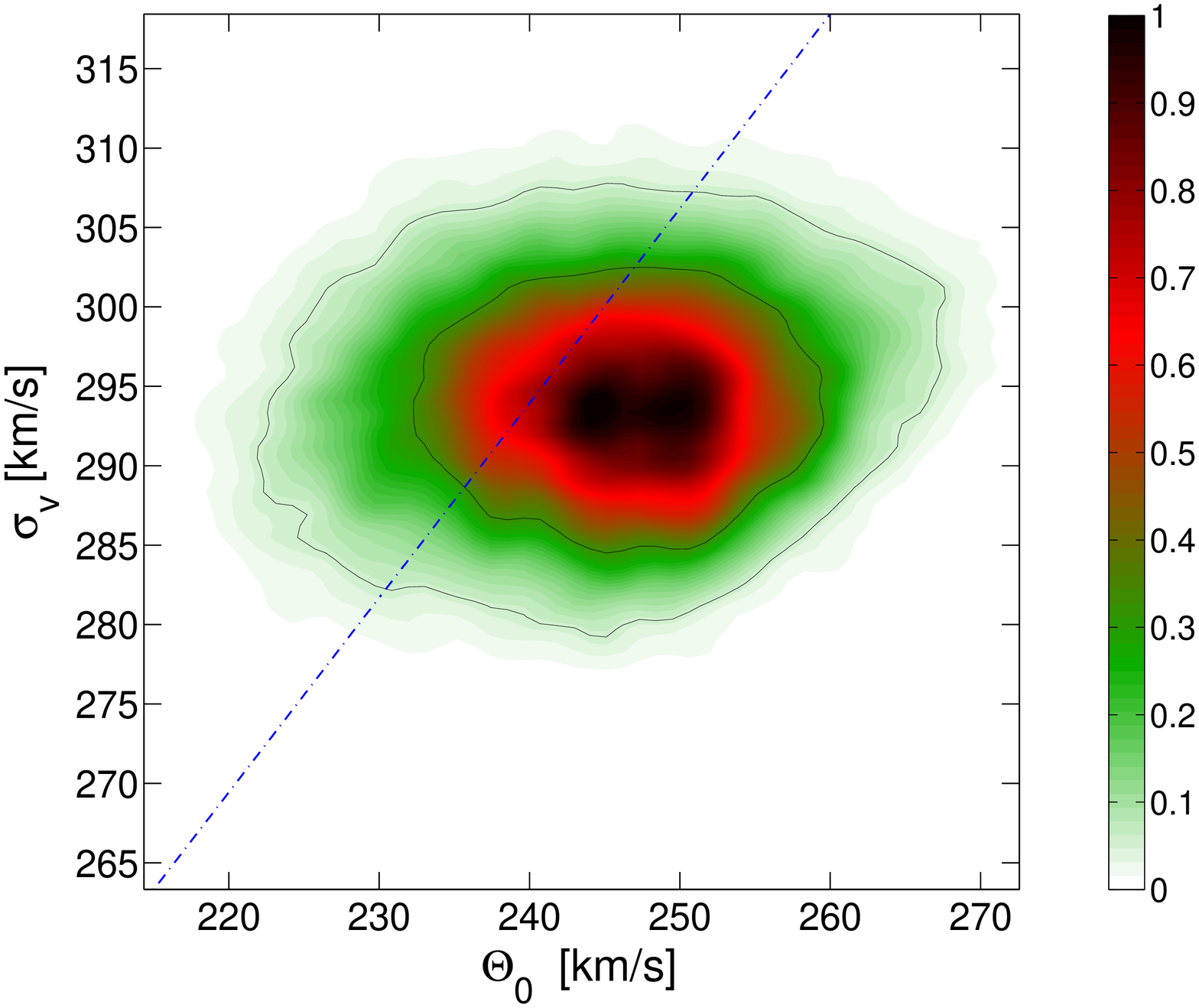}
\includegraphics[width=50mm,height=50mm]{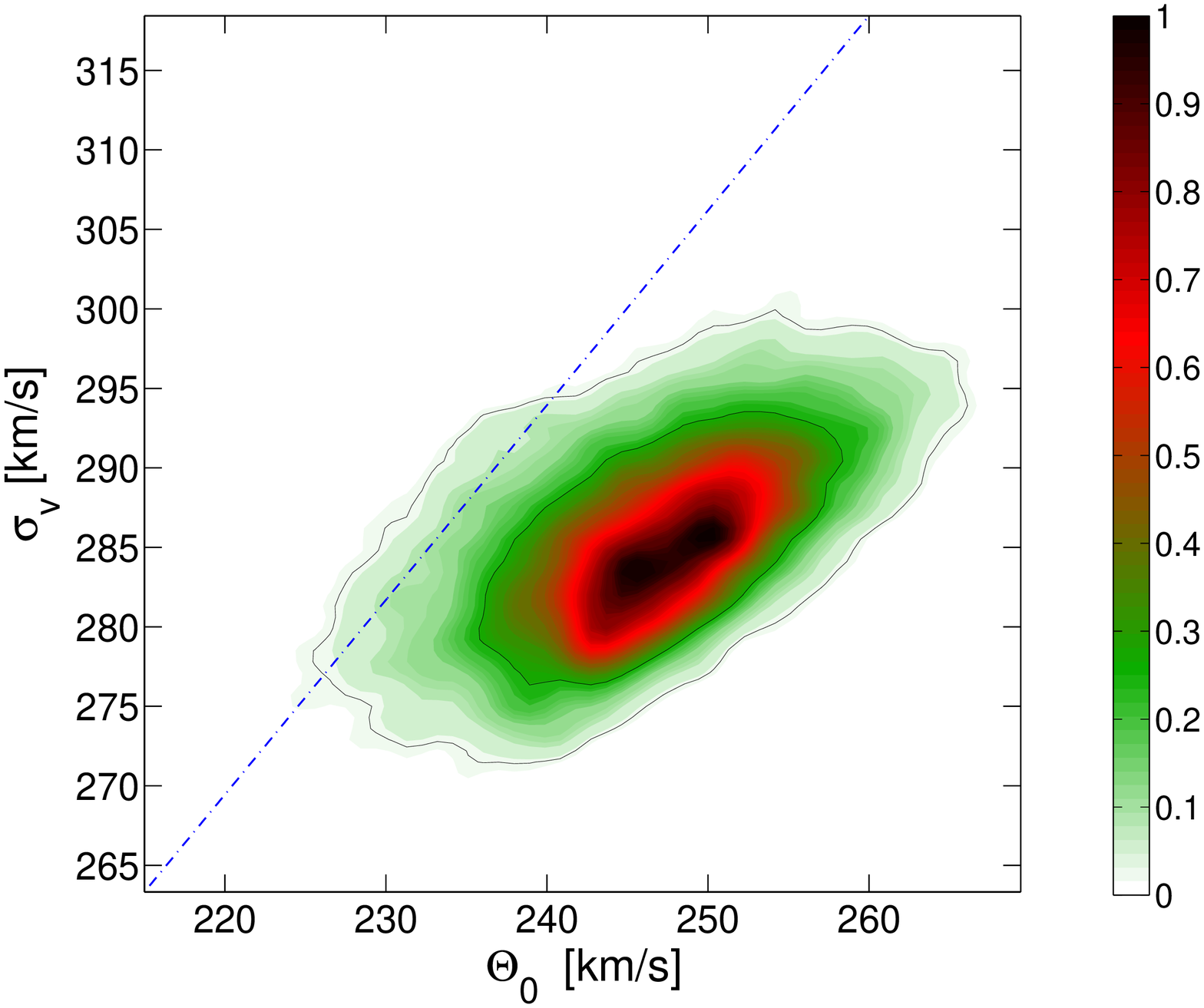}
\caption{2D marginal posterior pdf in the $(\Theta_0,\sigma_v)$ plane. Left panel: NFW profile. Central panel: Burkert profile. Right panel: Einasto profile. The blue (dashed) line represents the curve $\sigma_v=\sqrt{3/2}\,\Theta_0$}
\label{2D}
\end{figure}

Finally, in Fig.~\ref{2D} we show the two-dimensional marginal posterior pdf in the $(\Theta_0,\sigma_v)$ plane and compare the shape of this 2D-distribution with the straight line defined by  $\sigma_v = \sqrt{3/2} \,\Theta_0$. As one can see, the correlation pattern between these parameters suggested by the data is different from the one described by this linear relation. Indeed, the eigenvalues of the covariance matrix for the parameters $(\Theta_0,\sigma_v)$ point towards directions which are different from the one identified by the relation $\sigma_v = \sqrt{3/2} \,\Theta_0$. 

\subsection{Differential rate}
As a first application of our results on the DM phase-space density, we now discuss how uncertainties on this quantity affect the expectations for the differential event rate at a direct detection experiment. The differential event rate can be calculated using Eq.~(\ref{eq:ddrate3}) and depends on the Galactic model parameters through the phase-space density. It is however also a function of the observed recoil energy $E$ which enters in a convolution integral which involves the form factor and in the expression for $\vmin$. Similarly to what we did for the phase-space density, we now discretize the variable $E$ and study a set of quantities defined by $\zeta_k=\frac{dR}{dE}\left(E_k\right)$, where $E_k$ is a generic point in the observed energy discretization. Thus, to each energy $E_k$ we can now associate a marginal posterior pdf. The associated credibility intervals define the ``signal bands'' which incorporate the astrophysical uncertainties.

\begin{figure}
\includegraphics[width=50mm,height=50mm]{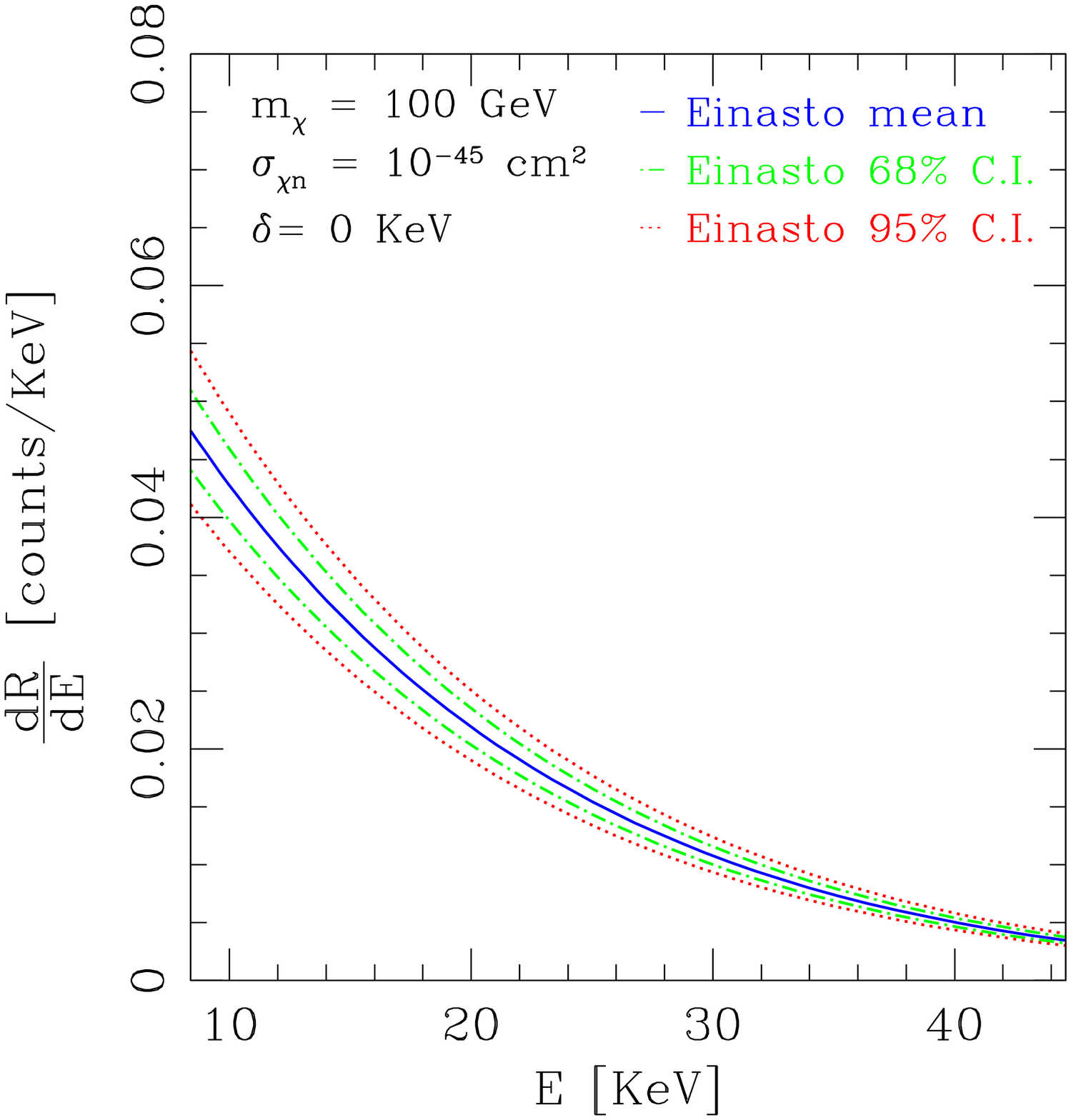}
\includegraphics[width=50mm,height=50mm]{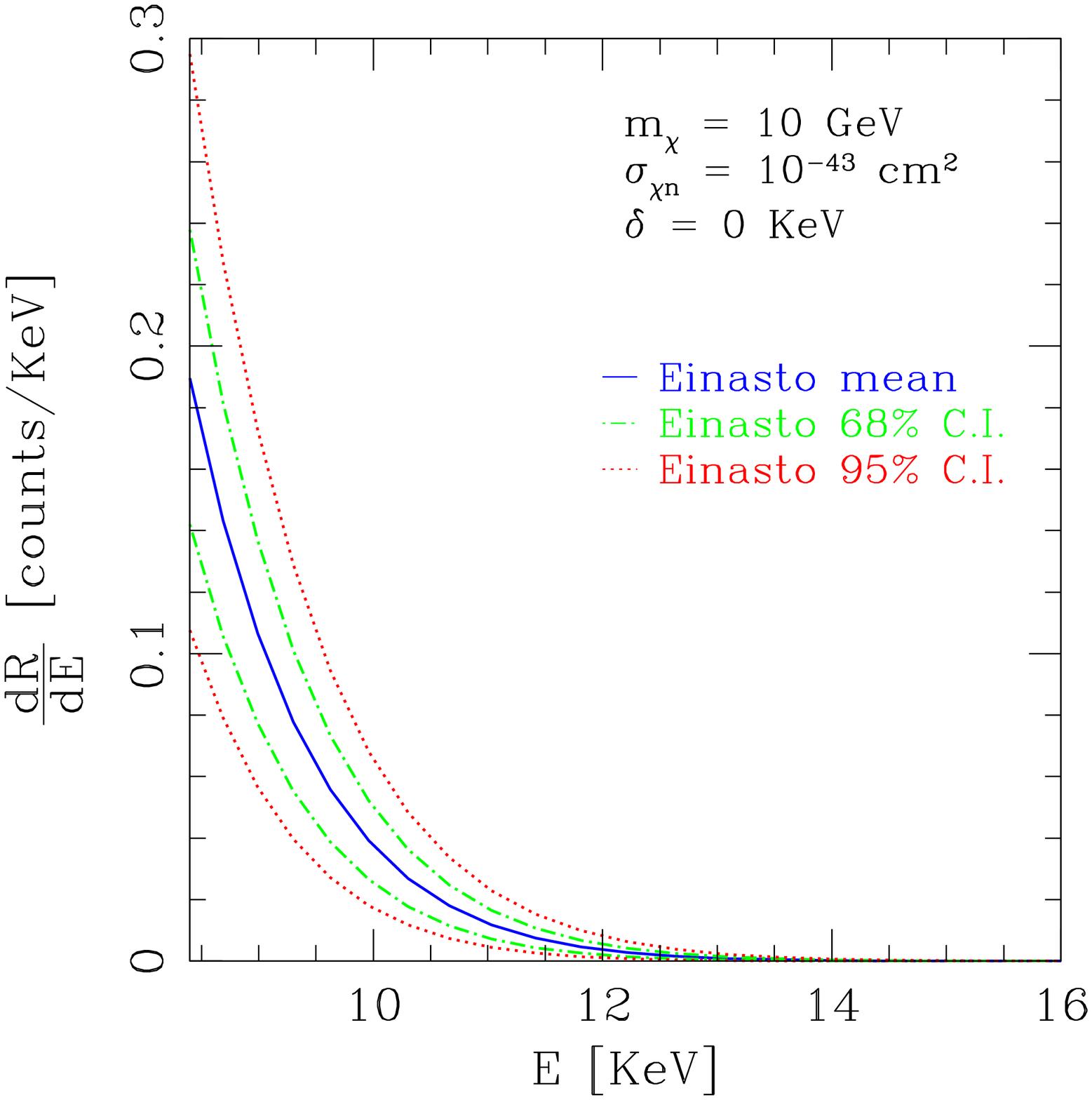}
\includegraphics[width=50mm,height=50mm]{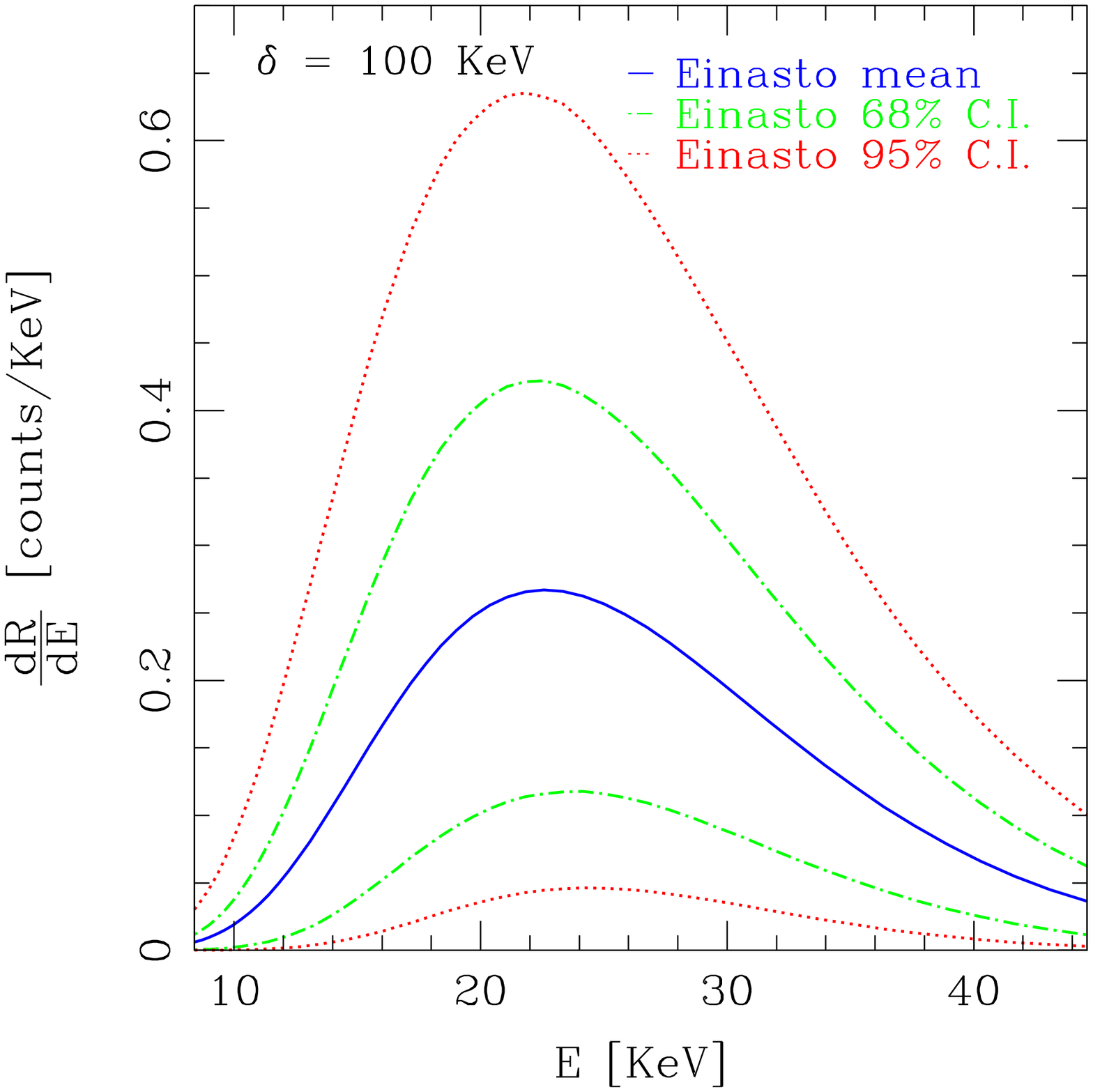}
\caption{MCMC differential event rate for the Einasto profile. Left panel: means and credibility intervals for a 100 GeV WIMP elastically scattering. Central panel: means and credibility intervals for a 10 GeV WIMP elastically scattering. Right panel: means and credibility intervals for a 45 GeV WIMP inelastically scattering with $\delta=100$ KeV.} 
\label{signals}
\end{figure}

In this section we present the results that we obtained following this approach for three illustrative cases: a 100 GeV WIMP (elastic scattering), a 10 GeV WIMP (elastic scattering) and, finally, a 45 GeV WIMP (inelastic scattering). The corresponding results are shown for the case of an Einasto profile in Fig.~\ref{signals}. In these plots a given credibility interval is the region between two curves of the same type. 

From these plots one can appreciate that for light candidates (Fig.~\ref{signals}, central panel), and in general in the case of an inelastic scattering (Fig.~\ref{signals}, right panel), astrophysical uncertainties can be extremely important and completely spoil the accuracy of the predicted event rate. This phenomenon is related to the fact that such candidates - light or inelastic - probe only the high velocity tail of the phase-space density. Therefore, a few per cent change in $\vesc$, for instance, can modify the integral over the phase-space density in Eq.~(\ref{eq:ddrate2}) by a factor $\gtrsim \mathcal{O}(1)$. Instead, heavier candidates scattering off the target nuclei elastically are less sensitive to astrophysical uncertainties because these are sensitive to a much larger portion of the phase-space density. Thus, in this case, the uncertainties on the differential rate - which are due to the uncertainties on the total area subtended by the phase-space density - are much less pronounced and are of the order of $\sim 20\%$.   

\begin{figure}
\includegraphics[width=50mm,height=50mm]{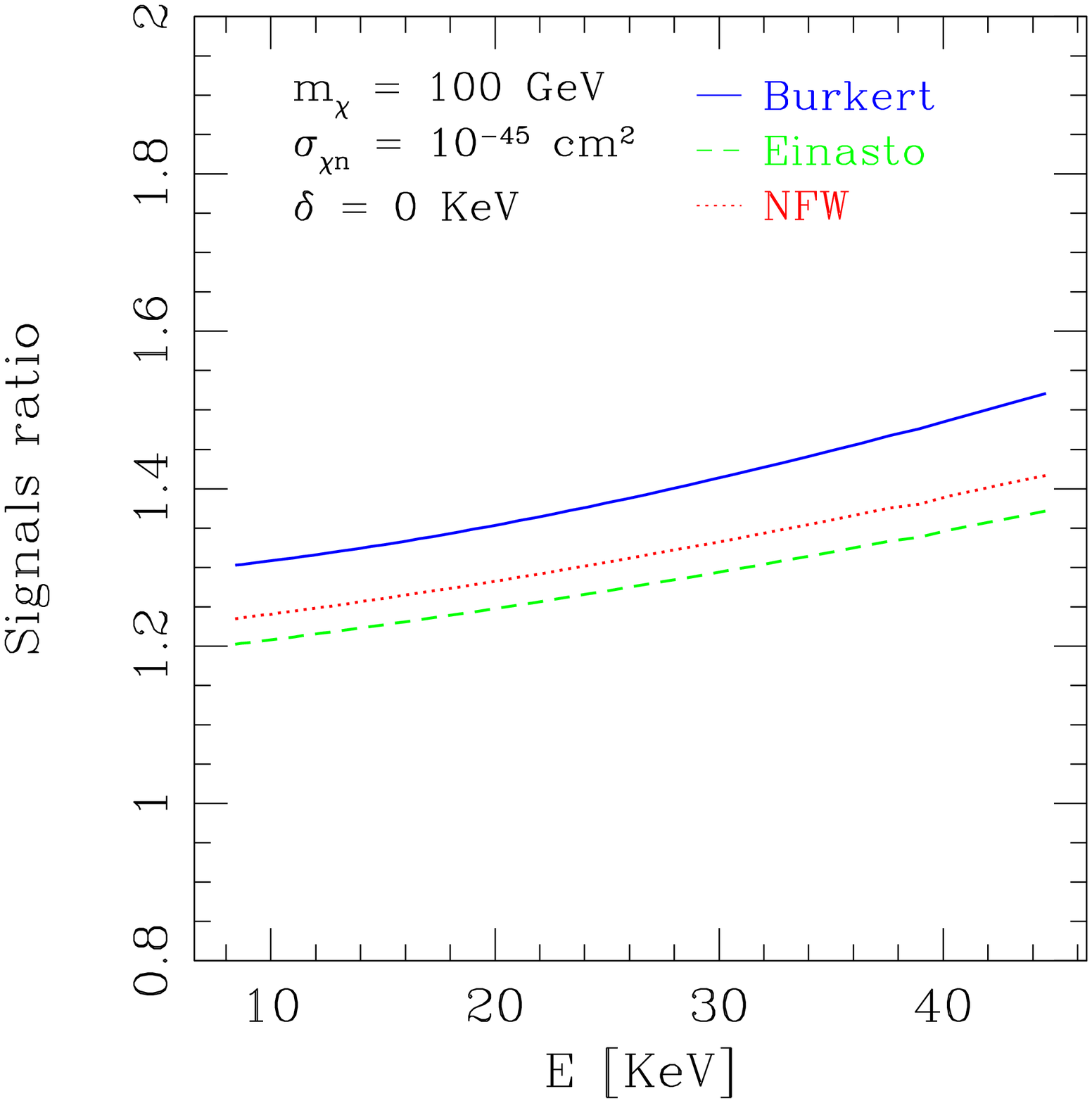}
\includegraphics[width=50mm,height=50mm]{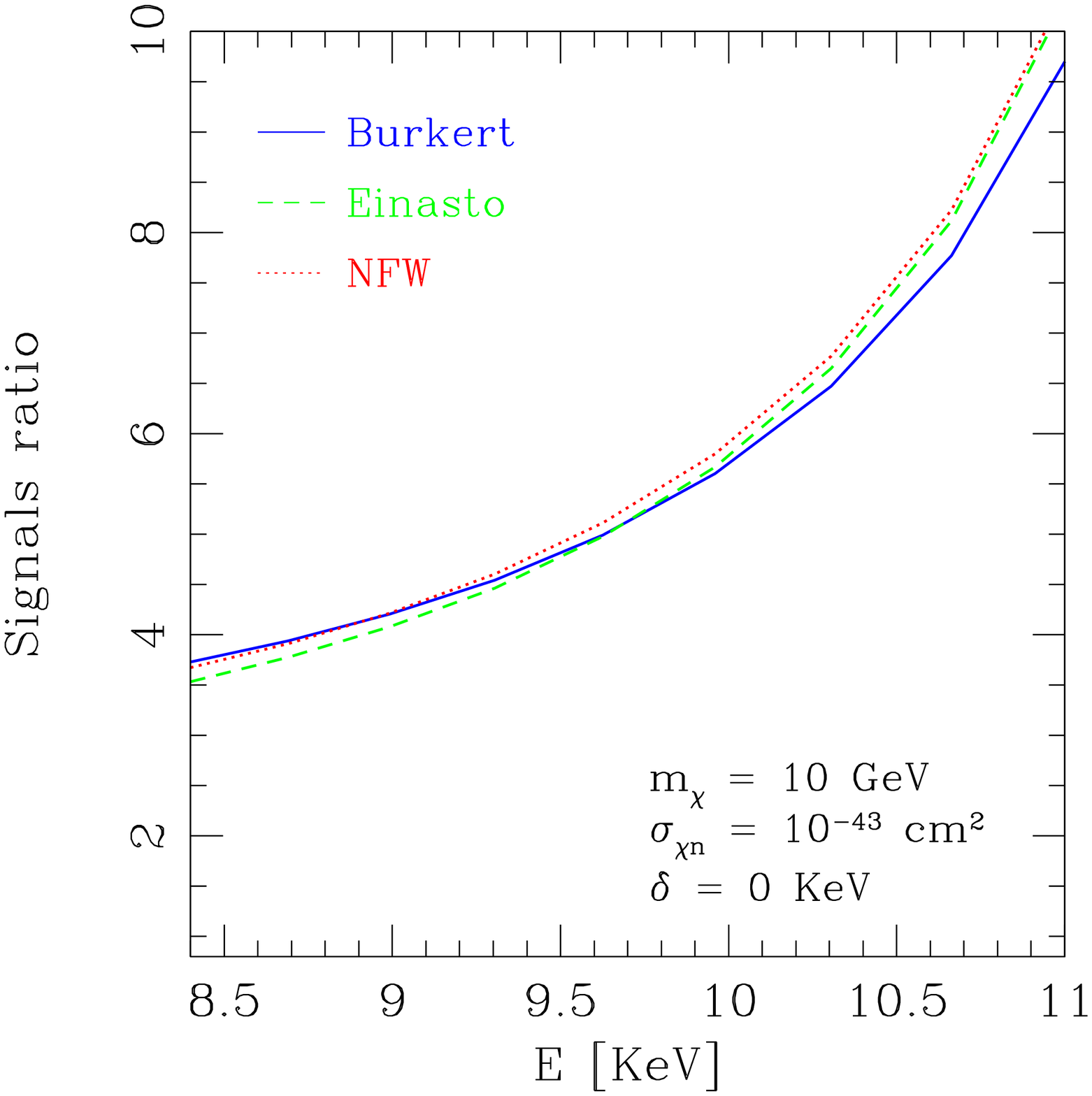}
\includegraphics[width=50mm,height=50mm]{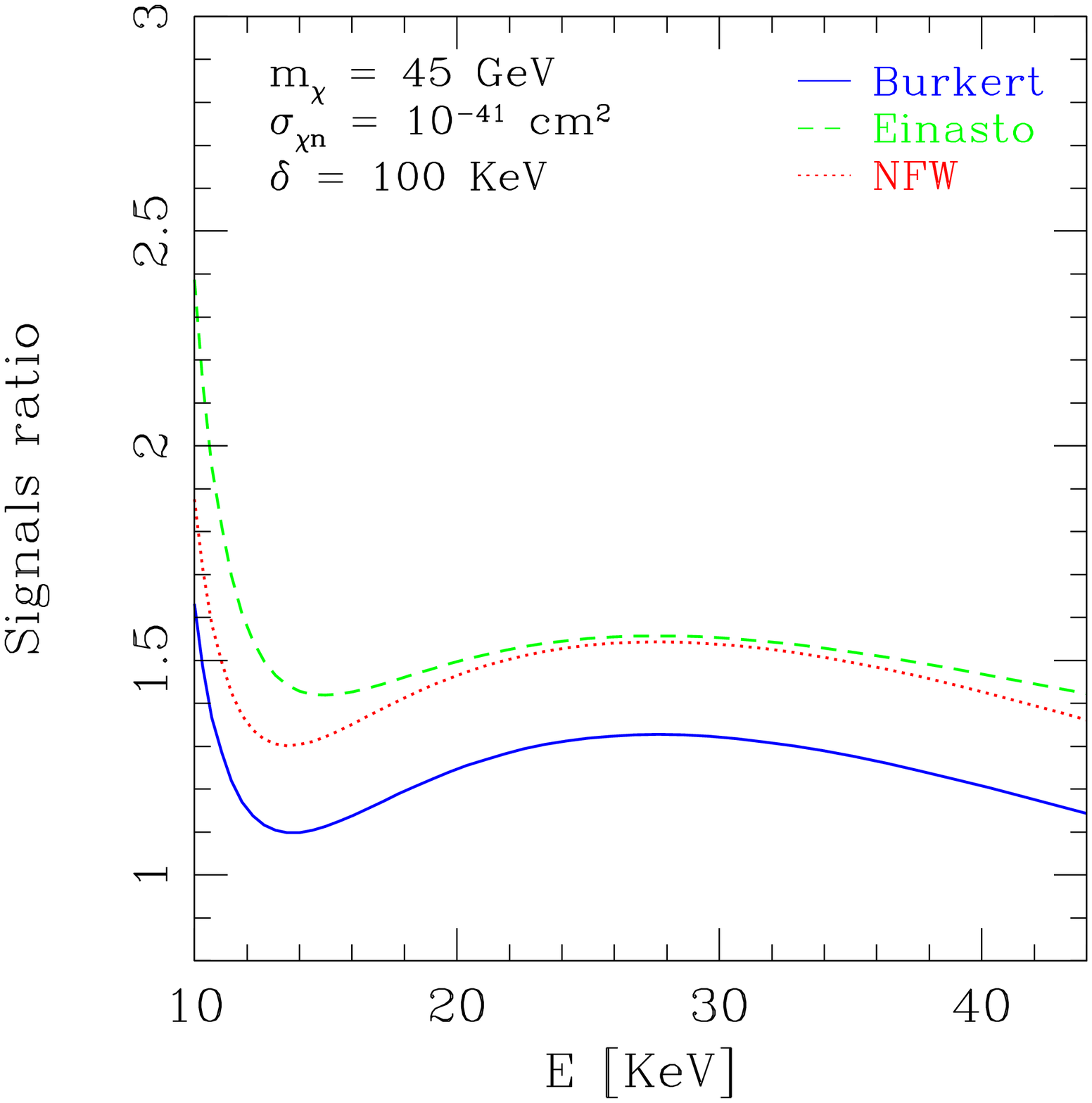}
\caption{Ratios between the MCMC differential event rates and the ones obtained in the Maxwell-Boltzmann approximation with $\Theta_0=220$ km/s, $\vesc=544$ km/s and $\rho_{\chi}=$ 0.3 GeV/cm$^3$. Left panel: 100 GeV WIMP elastically scattering. Central panel: 10 GeV WIMP elastically scattering. Right panel: 45 GeV WIMP inelastically scattering with $\delta=100$ KeV.}
\label{RatioSignals}
\end{figure}

In Fig.~\ref{RatioSignals} we compare the differential rates obtained assuming different DM profiles both for the elastic and for the inelastic scenario. In these plots all the curves have been normalized to the differential rate computed assuming a Maxwell-Boltzmann distribution with $\Theta_0=220$ km/s, $\vesc=544$ km/s and $\rho_{\chi}=$ 0.3 GeV/cm$^3$. In the case of a 100 GeV WIMP the curves shown in the plot differ mostly because the three corresponding profiles have different local DM densities. In the light and inelastic WIMP scenario, instead, the distance between the three curves also depend on the different high velocity behavior of the phase-space densities corresponding to the considered profiles. In all cases the Maxwell-Boltzmann approximation leads to lower differential rates.  

\subsection{Xenon100 exclusion limit}
Finally, we study how astrophysical uncertainties affect the estimation of the exclusion limit in the scattering cross section - mass plane. In this section we focus on the recent Xenon100 data, implementing a method based on the Poisson statistics, as explained in section \ref{ddrate}. We expect, however, similar conclusions, {\it i.e} similar credibility intervals, for different experiments analyzed with more refined statistical tools.

\begin{figure}
\begin{center}
\includegraphics[width=75mm,height=75mm]{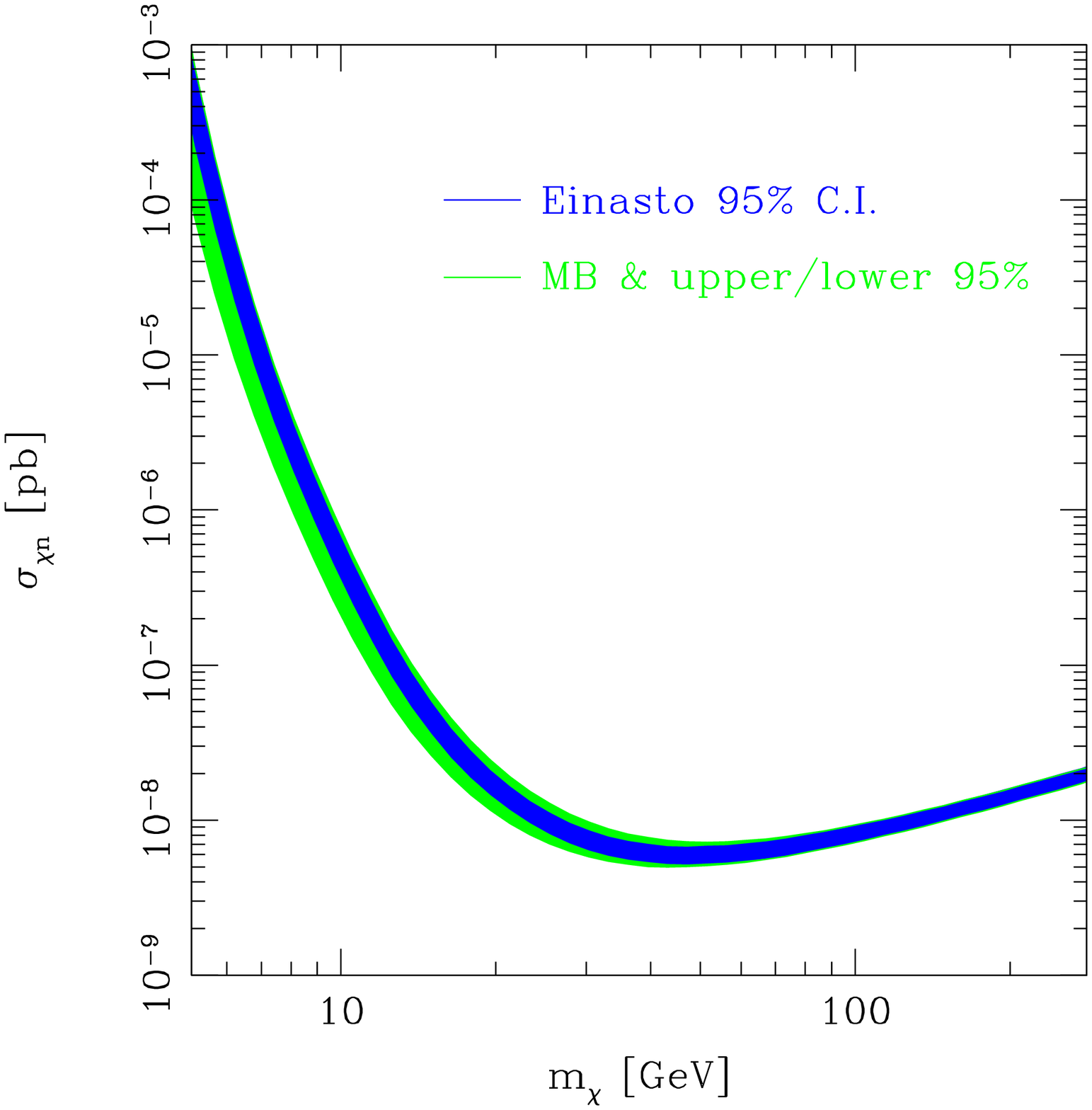}
\includegraphics[width=75mm,height=75mm]{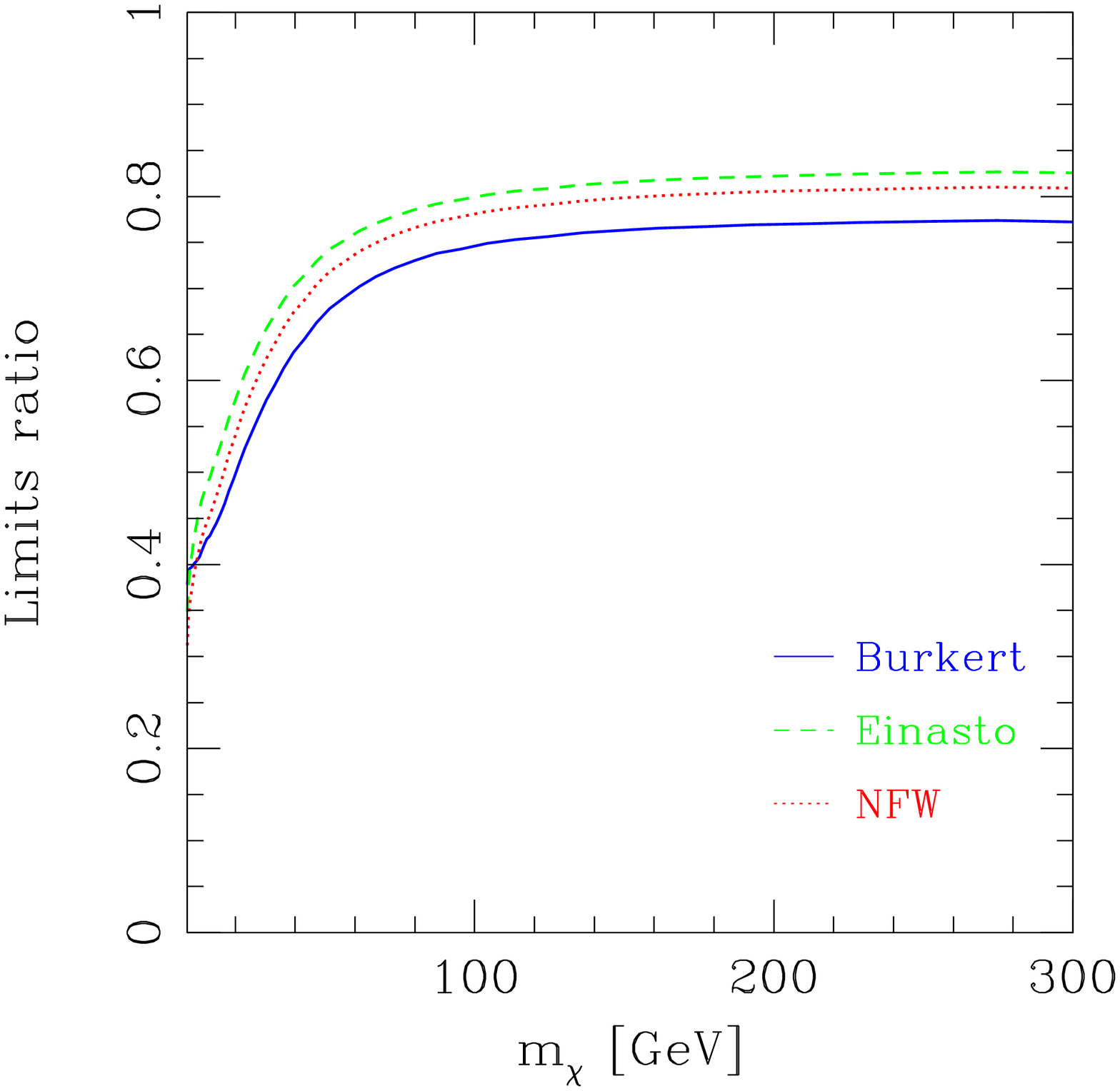}
\end{center}
\caption{Left panel: The blue (dark) band is the MCMC exclusion limit associated with the Einasto profile (95 \% C.I.). The green (light) band, instead, is the region between the exclusion limits obtained in the Maxwell-Boltzmann approximation with respectively $\Theta_0=229$ km/s, $\vesc=523$ km/s and $\rho_{\chi}=0.33$ GeV/cm$^3$ , and $\Theta_0=261$ km/s, $\vesc=589$ km/s and $\rho_{\chi}=0.45$ GeV/cm$^3$. Right panel: Ratios between the MCMC exclusion limits (three profiles) and the one obtained in the Maxwell-Boltzmann approximation with $\Theta_0=220$ km/s, $\vesc=544$ km/s and $\rho_{\chi}=$ 0.3 GeV/cm$^3$.} 
\label{limits}
\end{figure}

In the case of the exclusion limit, we discretize the mass $M_{\chi}$ of the DM candidate and study the quantities $\zeta_k=\bar{\sigma}(M_{\chi}^k)$, where $M_{\chi}^k$ is a point in the discretization of the WIMP mass, and $\bar{\sigma}$ is the upper bound defined in Eq.~(\ref{limit}). From the marginal posterior pdf's of the quantities $\bar{\sigma}(M_{\chi}^k)$ we derived a mean exclusion limit and the corresponding ``exclusion bands'' calculated from the credibility intervals associated with the different $\bar{\sigma}(M_{\chi}^k)$. The results are shown in Fig.~\ref{limits} (left panel), where we focus on the case of an Einasto profile. Together with the 95$\%$ MCMC credibility intervals we also plot for comparison the exclusion limits which one finds in the Maxwell-Boltzmann approximation with Galactic parameters set at their upper and lower 95$\%$ C.I. values (see Table \ref{tab_all}).      
In agreement with our results for the differential events rates, uncertainties at small masses are rather important while at larger masses they are less strong (in this respect the logarithmic scale in the plot might be misleading). 

In the right panel of Fig.\ref{limits} we show the ratios between the MCMC exclusion limits computed for the Einasto, NFW and Burkert profiles, and the one obtained in the Maxwell-Boltzmann approximation with $\Theta_0=220$ km/s, $\vesc=544$ km/s and $\rho_{\chi}=$ 0.3 GeV/cm$^3$. As for the differential rate, the curves differ because the three profiles are charcterized by different local densities and velocity distributions, being the tails of such distributions particularly relevant for low mass WIMPs. 

Fig.\ref{limits} also shows that in all cases the Maxwell-Boltzmann approximation leads to upper bounds for the scattering cross section less strong than in the MCMC computation, in particular at low masses where the Maxwell-Boltzmann distribution is artificially truncated at an escape velocity $\vesc=544$ km/s.      

\begin{figure}
\begin{center}
\includegraphics[width=75mm,height=75mm]{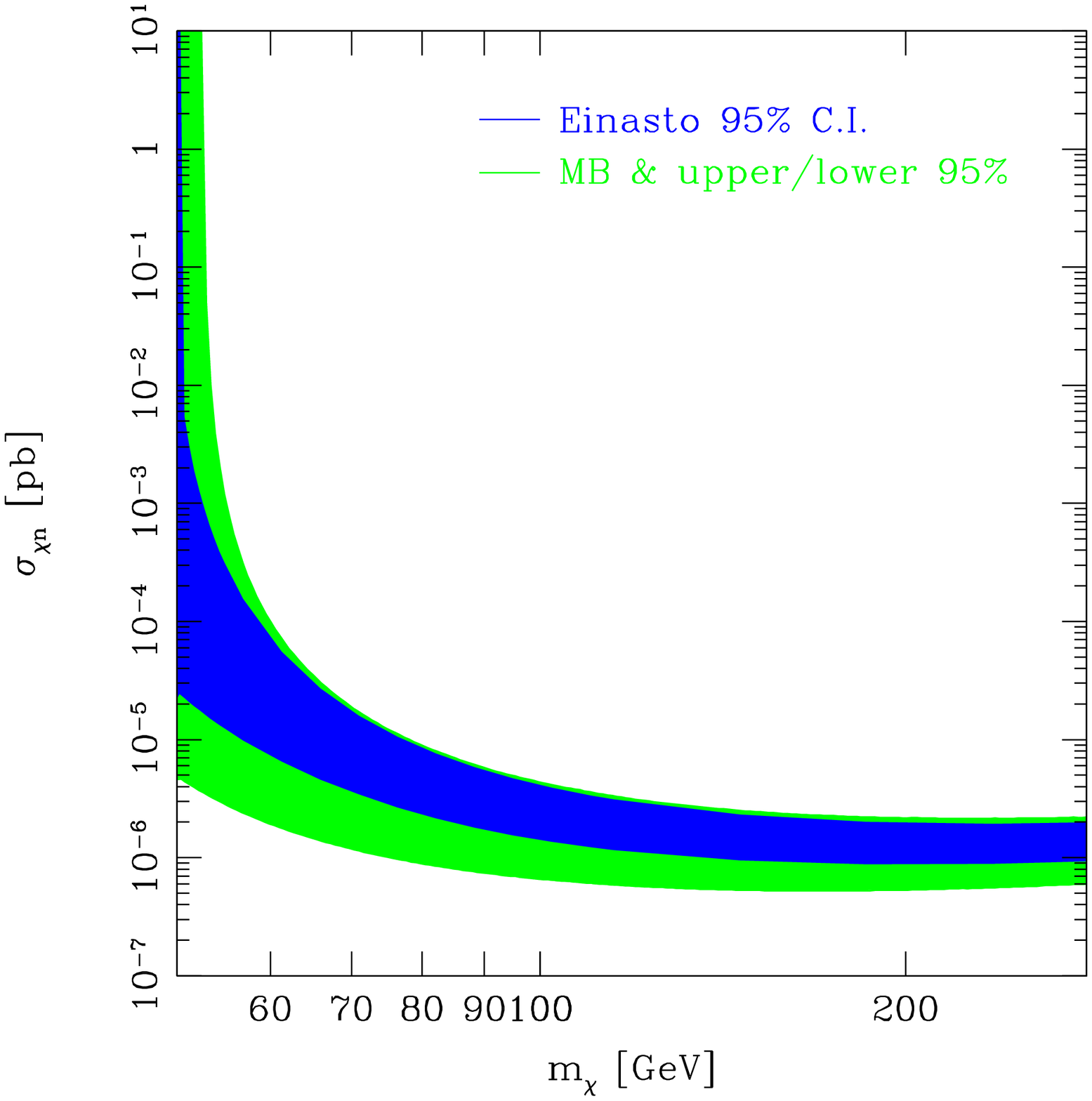}
\includegraphics[width=75mm,height=75mm]{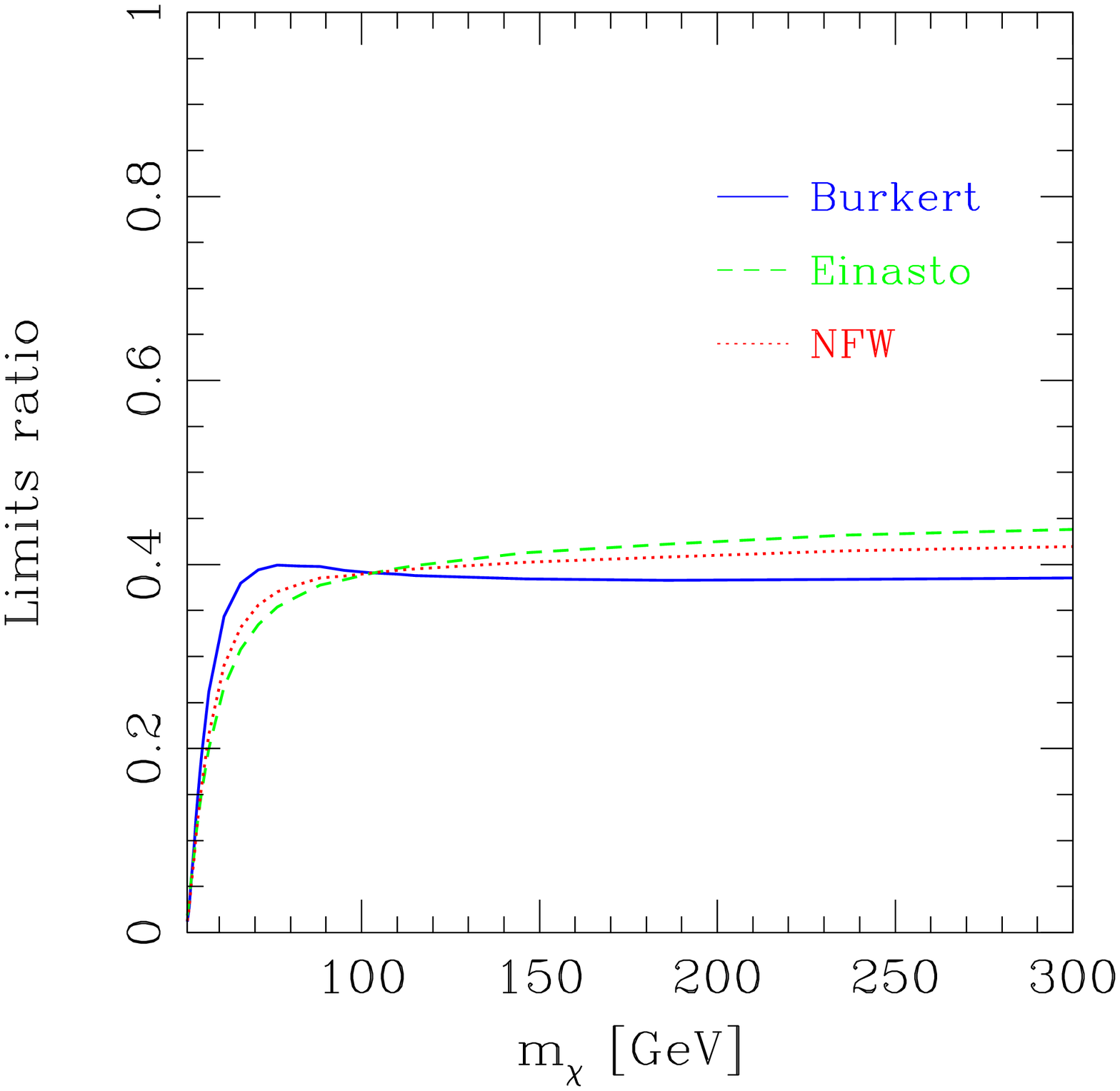}
\end{center}
\caption{As in Fig. \ref{limits} but with $\delta$=120 KeV.}
\label{limits-iDM}
\end{figure}

We also studied how astrophysical uncertainties affect the exclusion limit in the scattering cross section - mass plane for the inelastic DM scenario. The results obtained assuming an Einasto profile are shown in Fig.\ref{limits-iDM} (left panel). As for the case of DM elastically scattering off the detector nuclei, together with the 95$\%$ MCMC credibility intervals we also plot the results derived in the Maxwell-Boltzmann approximation with Galactic parameters set at their upper and lower 95$\%$ C.I. values (see Table \ref{tab_all}). As expected from our results on the differential rates, in the inelastic DM scenario uncertainties on the exclusion limit are considerably larger than in the case in which the DM - nucleus interaction occurs elastically. Finally, in the right panel of Fig.\ref{limits-iDM}, we show the ratios between the MCMC inelastic exclusion limits obtained for the Einasto, NFW and Burkert profiles, and the one computed in the Maxwell-Boltzmann approximation with $\Theta_0=220$ km/s, $\vesc=544$ km/s and $\rho_{\chi}=$ 0.3 GeV/cm$^3$.  As already mentioned presenting the analogous results for the elastic DM scenario, the curves differ bacause of the different local densities and velocity distributions characterizing the three considerd DM profiles.

\section{Discussion and Conclusions}
\label{concl} 
We have presented a new determination of the local DM phase-space density based on the Eddington's inversion formula. In this approach, assuming an isotropic velocity distribution, and in the limit of spherically symmetric DM density profile $\rho_h(r)$ and gravitational potential for the Galaxy $\Phi(r)$, there is an one-to-one correspondence between $\rho_h$ and the DM phase-space density.

Implementing, within a Bayesian framework, a broad and variegated sample of dynamical constraints for the Milky Way, and sampling through a MCMC algorithm the posterior pdf of the Galactic models parameters, we computed the local DM phase-space density. We have then applied this result to study a few illustrative cases of direct detection rates, and have discussed the impact on the exclusion limit one can infer from the latest data release from the Xenon experiment; the extension of our analysis to other DM models or to other experimental configurations is straightforward, and similar pattern are expected. The advantage of our approach is that it allows to keep a close inspection on the correlation among the astrophysical quantities setting the phase-space density of DM particles (and hence in turn the direct detection signals) and propagate uncertainties in a consistent way.  

We studied separately three cases corresponding to three possible choices for the DM density profile, namely the NFW, Burkert and Einasto profiles. We have found that, on average, the velocity dispersion, which
characterize the width in the velocity distribution, tends to be  larger for the Burkert profile than for the NFW or the Einasto. Studying two-dimensional marginal posterior probability in the plane velocity dispersion versus circular velocity, we have verified that for any of the three distributions these two quantities do not follow the linear correlation pattern suggested in the isothermal sphere model, and adopted in most analyses of direct detection rates.  We have computed the distributions of the escape velocities in our samples and found mean values which are consistent with the mean value obtained from observations of high velocity stars by the RAVE Survey; on the other hand, we have found that the shape of the high velocity tails in our distributions is sensibly different with respect to what you find assuming a Maxwell-Boltzmann distribution truncated at a given value of the escape velocity.

Concerning the DM direct detection signals, astrophysical uncertainties have the largest impact on models for which the scattering rate depends mainly on the tails of the phase-space density, namely for light WIMPs or DM candidates inelastically scattering, given that in such cases the integral of the phase-space density  
can be considerably different for different Galactic models. We have also compared our result with those obtained with a Maxwell-Boltzmann 
defined by setting circular velocity, escape velocity and local halo density to the "standard" reference values
adopted in most direct detection studies, respectively, $\Theta_0=220$ km/s, $\vesc=544$ km/s and $\rho_{\chi}=$ 0.3 GeV/cm$^3$. Regardless of the assumed DM profile, when adopting a self-consistent MCMC DM phase-space density, we have found that the differential rates are higher and the exclusion limits stronger than in the standard Maxwell-Boltzmann approximation.

Results in this paper are derived under the hypotheses of spherical symmetry and isotropy of the velocity distribution, two rather strong assumptions which may be (slightly) inaccurate for describing the DM distribution in the local neighborhood (see the note referring to this aspect in section \ref{df}). Having demonstrated here the power of the Bayesian approach in this analysis, we remark that such approach is general enough to be implemented within more general Galactic geometries, where extensions of the Eddington's inversion are required in order to relate the phase-space density to the parameter space of the underlying Galactic model; a closer inspection on this point will be carried out in future work.  

\acknowledgments
These results have been obtained making use of the bwGRiD Cluster (http://www.bw-grid.de), member of the German D-Grid initiative, funded by the Ministry for Education and Research (Bundesministerium f\"ur Bildung und Forschung) and the Ministry for Science, Research and Arts Baden-W\"urttemberg (Ministerium f\"ur Wissenschaft, Forschung und Kunst Baden-W\"urttemberg).

\appendix
\section{Approximate velocity distributions}

\begin{figure}
\includegraphics[width=50mm,height=50mm]{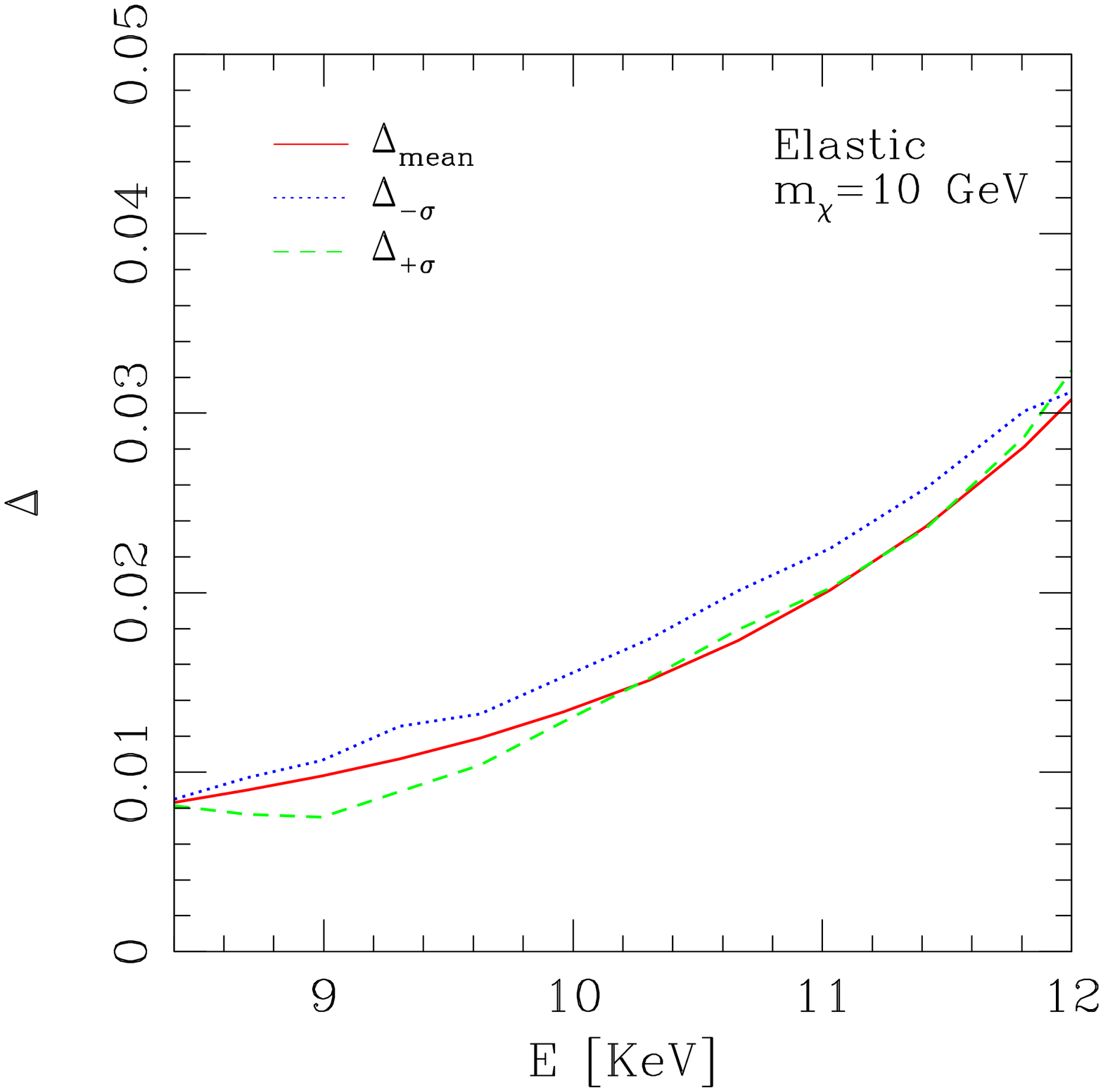}
\includegraphics[width=50mm,height=50mm]{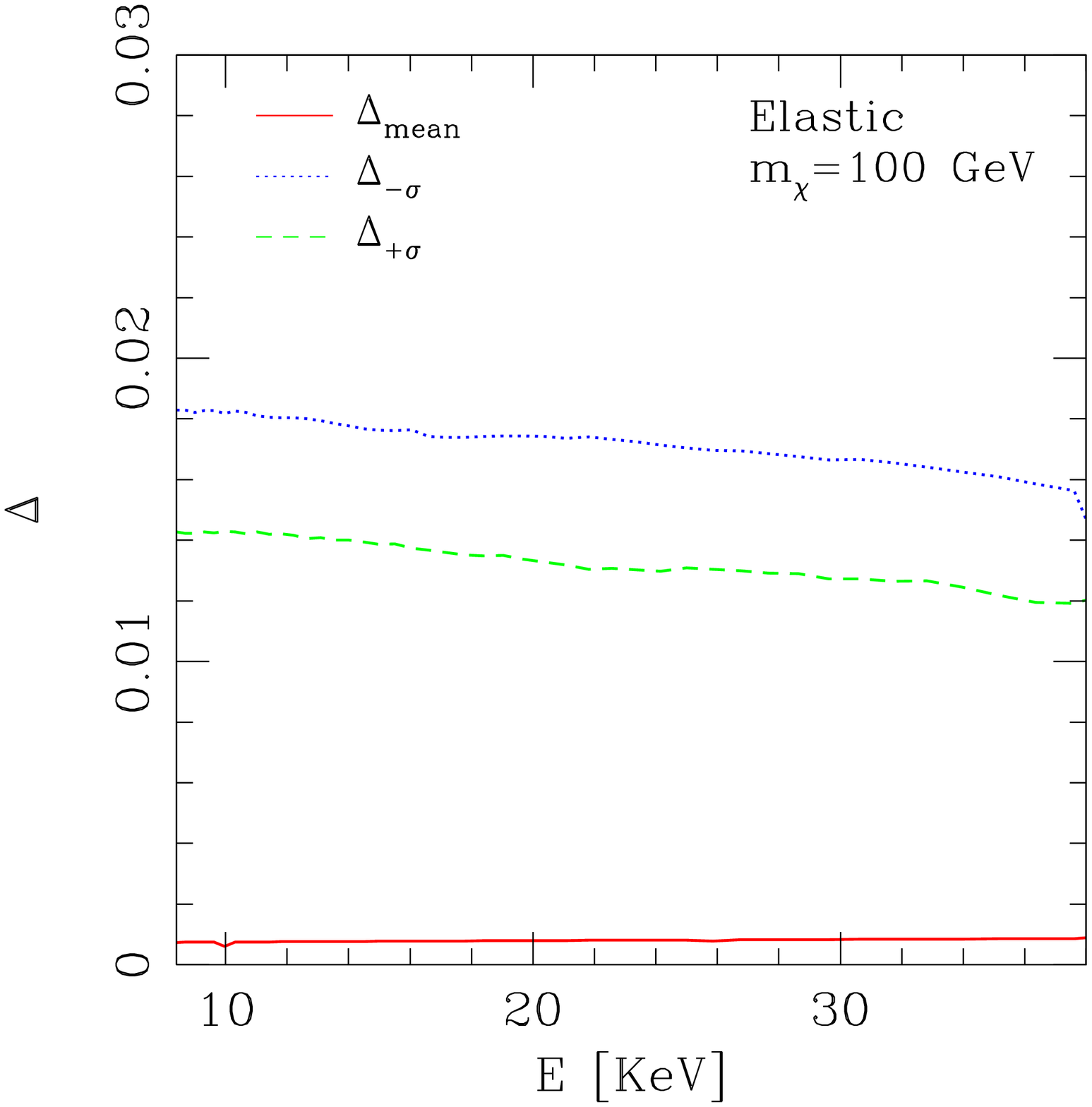}
\includegraphics[width=50mm,height=50mm]{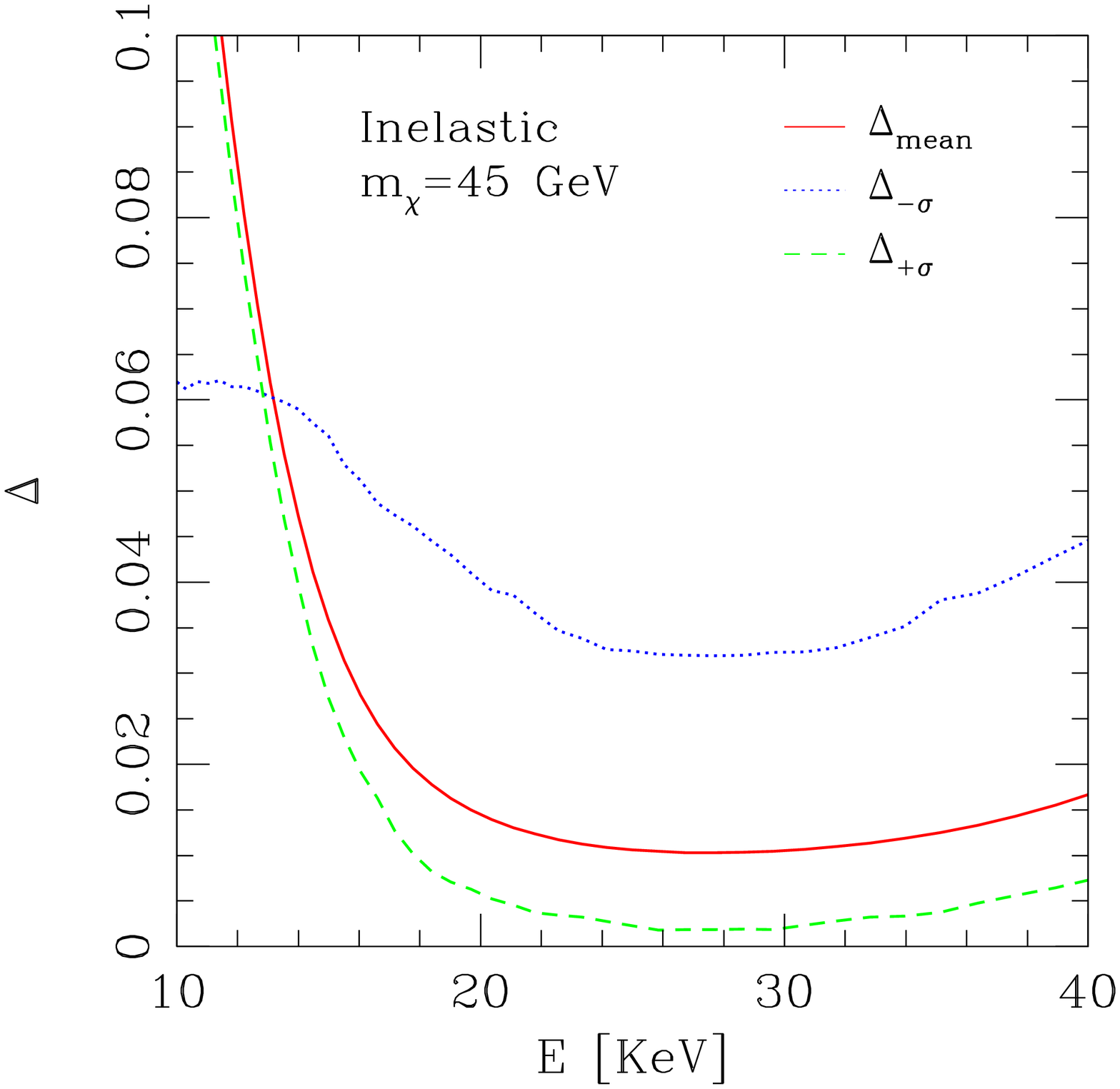}
\caption{MCMC signals vs the signals computed using the MCMC phase-space densities. We plot Eqs.~(\ref{eq:Delta}) for a 10 GeV elastic WIMP (left panel), a 100 GeV elastic WIMP (central panel) and an inelastic candidate ($\delta=100$ KeV) of 45 GeV (right panel). Relative errors are of a few $\%$ in regions where the signal is significantly different from zero. Results have been presented for the NFW case.}
\label{Delta}
\end{figure}

The results presented in this paper rely on the MCMC chains generated during our sampling of the posterior pdf and have been derived for specific WIMPs masses, cross sections or experimental configurations. To apply our phase-space densities to the study of different WIMPs or experimental setups, one can use the following approximation
\beq
\langle\frac{dR}{dQ}\rangle \simeq \frac{\sigma_{\chi n}}{2\,M_{\chi}M_n^2} A^2 {\mathcal F}_N^2(Q)
  \int_{u > \vmin} du\; \langle g_\chi \rangle \equiv \hat{\frac{dR}{dQ}}\,.
\label{approx}
\eeq
In Eqs.~(\ref{approx}) we are approximating the mean MCMC differential rate with the differential rate calculated using the mean MCMC phase-space density. This approximation is very good for $\sim$100 GeV WIMPs and is accurate at the per cent level for light WIMPs or inelastic DM candidates in the range of energies where the corresponding signals are significantly different from zero (see Fig.~\ref{Delta}). Error bars on direct detection signals can be then estimated calculating an upper limit $\sigma^{(+)}_{R}$ and a lower limit $\sigma_{R}^{(-)}$ analogously:
\beq
\sigma^{(\pm)}_{R} \simeq \frac{\sigma_{\chi n}}{2\,M_{\chi}M_n^2} A^2 F_N^2(Q)
  \int_{u > \vmin} du\; \left(\langle g_\chi \rangle \pm \sigma_{g_{\chi}}\right) \equiv \hat{\sigma}^{(\pm)}_{R}\,,
\label{approx2}
\eeq
where $\sigma_{g_{\chi}}$ can be computed as in Eqs.~(\ref{ev}).
In Fig.~\ref{Delta} we explicitly checked the validity of this approximation by studying how the quantities\footnote{ Indeed, in Fig.~\ref{Delta} we also consider finite resolution effects according to Eq.~(\ref{eq:ddrate3}).} 
\beq
\Delta = 2\times\frac{|\langle\frac{dR}{dQ}\rangle - \hat{\frac{dR}{dQ}}|}{\langle\frac{dR}{dQ}\rangle + \hat{\frac{dR}{dQ}}} \,;\qquad \qquad
\Delta_{\pm\sigma} = 2\times\frac{|\sigma^{(\pm)}_{R} - \hat{\sigma}^{(\pm)}_{R}|}{\sigma^{(\pm)}_{R} + \hat{\sigma}^{(\pm)}_{R}}
\label{eq:Delta}
\eeq
depend from the recoil energy. As already mentioned, when the signal is non-negligible, the relative errors $\Delta$ and $\Delta_{\pm}$ are of a few per cent or less. Eqs.~(\ref{approx}) and (\ref{approx2}) should be therefore useful for a quick estimation of DM signals which depend from the local DM phase-space density. Files including time averaged $\langle g_\chi \rangle$, $\langle g_\chi \rangle \pm \sigma_{g_{\chi}}$ and $\langle g_\chi \rangle \pm 2\sigma_{g_{\chi}}$for the three considered DM profiles are provided with the electronic version of the paper.  

\section{Data files provided with the paper}

With the electronic version of the paper we provide fifteen data files. They contain different tabulations of $g_{\chi}$ as a function of the velocity $u$ in the detector rest frame as calculated by using Eq.~(\ref{gfun}). There are five files for each profile considered in the paper, corresponding to $\langle g_{\chi} \rangle$, $\langle g_\chi \rangle \pm \sigma_{g_{\chi}}$ and $\langle g_\chi \rangle \pm 2\sigma_{g_{\chi}}$ respectively. Each file has a name of the form: 
\beq
\textrm{uDF$_{-}$halotype$_{-}$curvetype.dat}
\nonumber
\eeq
with halotype=nfw,burkert,einasto and curvetype=mean, 1splus, 1sminus, 2splus, 2sminus. 1splus and 1sminus correspond to $\langle g_\chi \rangle + \sigma_{g_{\chi}}$ and $\langle g_\chi \rangle - \sigma_{g_{\chi}}$ respectively, while 2splus and 2sminus are instead associated with $\langle g_\chi \rangle + 2\sigma_{g_{\chi}}$ and $\langle g_\chi \rangle - 2\sigma_{g_{\chi}}$. The velocity is in km/s while $g_{\chi}$ is in (GeV/cm$^3$)(s$^2$/km$^2$) and has been time averaged. The format of the files is the one required by the \ds\ function dshmuDFnum, which means that the first two lines are two empty headers and the data just start from the third line. 

\section{The role of the priors}
\label{axial}

\begin{figure}
\begin{center}
\includegraphics[width=10 cm,height=10 cm]{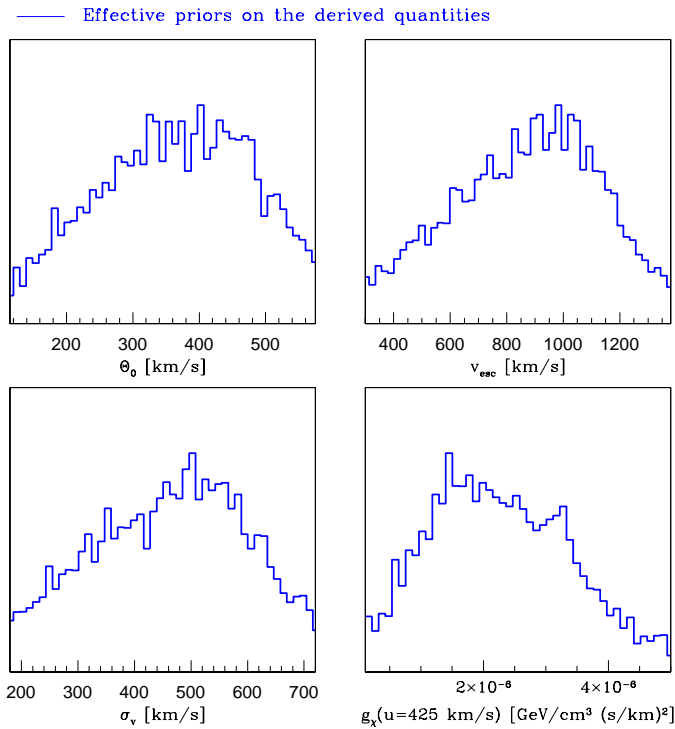}
\caption{Effective priors on the local circular velocity, escape velocity, velocity dispersion and the value of the function $g_{\chi}$ evaluated at the representative value of $420$ km s$^{-1}$.}
\label{priors}
\end{center}
\end{figure}

In this appendix we will focus on the impact of the choice of the priors on our conclusions. In the present analysis we sampled the posterior pdf using flat priors for the Galactic model parameters, however, this does not guarantee that the priors actually imposed on the derived quantities are also non-informative. To determine the effective priors used for the local circular velocity, escape velocity, velocity dispersion and the values of the function $g_{\chi}$, following \cite{Trotta:2008bp}, we sampled the posterior pdf for these quantities with a MCMC scan where no constraints were imposed on the underlying Galactic model parameters. The pdf found in this way provides an estimation of the effective priors used for the derived quantities. In Fig. \ref{priors} one can appreciate that, though the actual priors on these quantities are not ``rigorously flat'', the obtained results are not dominated by the priors as well, since a high probability is a priori assigned to a large range of possible values for the derived quantities.

%%%%%%%%%%%%%%%%%%%%%%%%%%%%%%%%%%%%%%%%

\end{document}